\documentclass[twocolumn,twocolappendix,iop,apj]{emulateapj}
\usepackage{times}

\usepackage{verbatim}
\usepackage[normalem]{ulem}
\usepackage[usenames, dvipsnames]{color}
\newcommand{\imos}[1]{#1}

\newcommand{\myemail}{Corresponding Mail}

\newcommand{\degree}{\ensuremath{^\circ }}

\newcommand{\fermilat}{\emph{Fermi}-LAT}

\newcommand{\hi}{H\,{\sc i}}
\newcommand{\hii}{H\,{\sc ii}}
\newcommand{\htwo}{H$_2$}
\newcommand{\galprop}{GALPROP}
\newcommand{\gray}{$\gamma$-ray}

\slugcomment{Draft: \today}

\shorttitle{Solution of heliospheric propagation}
\shortauthors{Boschini et al.}

\usepackage{amsmath} 

\begin{document}


\title{Solution of heliospheric propagation: unveiling the local interstellar spectra \\of cosmic ray species
}


\author{
M.~J.~Boschini\altaffilmark{1,2},  S.~{Della~Torre}\altaffilmark{1}, M.~Gervasi\altaffilmark{1,3}, D.~Grandi\altaffilmark{1},
G.~J\'{o}hannesson\altaffilmark{4,5}, M.~Kachelrie\ss\altaffilmark{6}, G.~{La~Vacca}\altaffilmark{1}, N.~Masi\altaffilmark{7},
I.~V.~Moskalenko\altaffilmark{8,9}, E.~Orlando\altaffilmark{8,9}, S.~S.~Ostapchenko\altaffilmark{10,11},
S.~Pensotti\altaffilmark{1,3}, T.~A.~Porter\altaffilmark{8}, L.~Quadrani\altaffilmark{7,12}, P.~G.~Rancoita\altaffilmark{1},
D.~Rozza\altaffilmark{1,3}, and M.~Tacconi\altaffilmark{1,3}
}
\email{\myemail}


\altaffiltext{1}{INFN, Milano-Bicocca, Milano, Italy}
\altaffiltext{2}{also CINECA, Segrate, Milano, Italy}
\altaffiltext{3}{also Physics Department, University of Milano-Bicocca, Milano, Italy}
\altaffiltext{4}{Science Institute, University of Iceland, Dunhaga 3, IS-107 Reykjavik, Iceland}
\altaffiltext{5}{also NORDITA,  Roslagstullsbacken 23, 106 91 Stockholm, Sweden}
\altaffiltext{6}{Institutt for fysikk, NTNU, 7491 Trondheim, Norway}
\altaffiltext{7}{INFN, Bologna, Italy}
\altaffiltext{8}{Hansen Experimental Physics Laboratory, Stanford University, Stanford, CA 94305}
\altaffiltext{9}{Kavli Institute for Particle Astrophysics and Cosmology, Stanford University, Stanford, CA 94305}
\altaffiltext{10}{Frankfurt Institute of Advanced Studies, Frankfurt, Germany}
\altaffiltext{11}{Skobeltsyn Institute of Nuclear Physics, Moscow State University, 119991 Moscow, Russia}
\altaffiltext{12}{also, Physics Department, University of Bologna, Bologna, Italy}


\begin{abstract}

Local interstellar spectra (LIS) for protons, helium and antiprotons are built using the most recent experimental results combined with the state-of-the-art models for propagation in the Galaxy and heliosphere. Two propagation packages, \galprop{} and HelMod, are combined to provide a single framework that is run to reproduce direct measurements of cosmic ray (CR) species at different modulation levels and at both polarities of the solar magnetic field. To do so in a self-consistent way, an iterative procedure was developed, where the GALPROP LIS output is fed into HelMod that provides modulated spectra for specific time periods of selected experiments to compare with the data; \imos{the HelMod parameters optimization is performed at this stage and looped back to adjust the LIS using the new GALPROP run.} The parameters were tuned with the maximum likelihood procedure using an extensive data set of proton spectra from 1997--2015. The proposed LIS accommodate both the low energy interstellar CR spectra measured by Voyager 1 and the high energy observations by BESS, Pamela, AMS-01, and AMS-02 made from the balloons and near-Earth payloads; it also accounts for Ulysses counting rate features measured out of the ecliptic plane. The found solution is in a good agreement with proton, helium, and antiproton data by AMS-02, BESS, and PAMELA in the whole energy range.

\end{abstract}



\section{Introduction} \label{Introduction} \label{Intr}
Considerable advances in astrophysics of CRs in recent years have become possible due
to superior instrumentation launched into space and to the top of the atmosphere.
The launch of Payload for Antimatter Matter Exploration and Light-nuclei Astrophysics \citep[PAMELA,][]{2007APh....27..296P} in 2006, followed by the {\it Fermi} Large Area Telescope \citep[\fermilat,][]{2009ApJ...697.1071A} in 2008, and the Alpha Magnetic Spectrometer--02 \citep[AMS--02,][]{2013PhRvL.110n1102A}
in 2011 signify the beginning of a new era in astrophysics. New materials and technologies employed by these space missions have enabled
measurements with unmatched precision, which allows for searches of subtle signatures of new phenomena in CR and \gray{} data.

These advances are built on solid results of earlier missions.
Understanding the origin of CRs, their acceleration mechanisms, main features of the interstellar propagation,
and CR source composition would be impossible without extraordinary efforts of the teams built around
Cosmic Ray Isotope Spectrometer (CRIS) onboard NASA's Advanced Composition Explorer~\citep[ACE,][]{1998SSRv...86..285S},
Super Trans-Iron Galactic Element Recorder~\citep[SuperTIGER,][]{2014ApJ...788...18B}, and other
(earlier) experiments such as ATIC, BESS, CAPRICE, CREAM, HEAO-3, HEAT, ISOMAX, TIGER, TRACER,
Ulysses, and others.
Launched in 1977 at the dawn of the space era, Voyager~1,~2 spacecrafts \citep{1977SSRv...21..355S} demonstrate
unbelievable scientific longevity providing unique data on the elemental spectra and composition
at the interstellar reaches of the Solar system, currently at 138 AU and 114 AU from the Sun, correspondingly.
Other high-expectations missions are recently launched (like the CALorimetric Electron Telescope -- \citealp[CALET,][]{2015JPhCS.632a2023A} -- and the Dark Matter Particle Explorer -- \citealp[DAMPE,][]{Azzarello2016378}) or are awaiting for launch~\citep[Cosmic-Ray Energetics and Mass investigation --
ISS-CREAM,][]{2014AdSpR..53.1451S}.

Indirect CR measurements are made by multi-wavelength observatories.
\fermilat{} is mapping the all-sky diffuse \gray{} emission, produced by CR interactions in the
interstellar medium (ISM), and near CR accelerators.
The International Gamma-Ray Astrophysics Laboratory~\citep[INTEGRAL,][]{2003A&A...411L...1W}, the
High-Altitude Water Cherenkov Observatory~\citep[HAWC,][]{2013APh....50...26A},
the High Energy Stereoscopic System~\citep[\imos{H.E.S.S.,}][]{2009ARA&A..47..523H}, Major
Atmospheric Gamma-ray Imaging Cherenkov Telescopes~\citep[MAGIC,][]{2016APh....72...76A}, and the
Very Energetic Radiation Imaging Telescope Array System~\citep[VERITAS,][]{2006APh....25..391H},
observe keV--TeV emissions produced by CR particles in various environments.
Construction of the first pre-production telescopes of the next generation Cherenkov Telescope Array \citep[CTA,][]{2013APh....43....3A} will begin in 2018.
High-resolution data in the microwave domain
are provided by the {\it Wilkinson} Microwave Anisotropy Probe \citep[WMAP,][]{2003ApJ...583....1B} and {\it Planck}
\citep{2010A&A...520A...1T}.

Our understanding of CR propagation in the Milky Way comes from a combination of observational data and a strong theoretical effort \imos{\citep[see][]{2007ARNPS..57..285S}}. Interpretation of many different kinds of data with a self-consistent model requires a state-of-the-art numerical tool that combines the latest information on the Galactic structure (distributions of gas, dust, radiation and magnetic fields) with the latest formalisms describing particle and nuclear cross sections and theoretical description of the processes in the ISM. This was realized about 20 years ago, when some of us started to develop the most advanced fully numerical CR propagation code, called \galprop{}\footnote{http://galprop.stanford.edu} \citep{1998ApJ...493..694M,1998ApJ...509..212S}. Over these years the project was widely recognized as a standard model for Galactic CR propagation and associated diffuse emissions (radio, X-rays, \gray{s}). \galprop{} uses information from astronomy, particle and nuclear physics to predict, in a self-consistent manner, CRs fluxes, \gray{s}, synchrotron emission and its polarization \citep[see][]{2007ARNPS..57..285S}, like a puzzle being assembled from the results of individual measurements in physics and astronomy spanning in energy coverage, types of instrumentation, and the nature of detected species and emissions. The project stimulated studies of the interstellar gas (\htwo, \hi{}, \hii{}), interstellar radiation and magnetic fields, and isotopic and particle production cross sections. These studies provide unique input datasets for the \galprop{} model.

An accurate description of propagation of CR particles through the heliosphere in the last $\sim$130 AU, that is a minuscule distance by the Galactic scale, was a considerable challenge until now. These last 0.0006 pc are so important because they provide a link between the predictions of the interstellar propagation models with the location where 99.9\% of all direct CR measurements are made. Even though, the heliospheric modulation affects only particles with small to medium energies below 30--50 GeV, this range includes the sub-GeV energies where the most precise measurements of CR isotopic composition are made. These low energy data are used to derive the parameters of interstellar propagation that are then extrapolated onto the whole Galaxy and all energies up to the multi-TeV region. Therefore, an improvement in the description of the heliospheric propagation has a global impact on our understanding of CR phenomena in the whole Galaxy, i.e., it is where many ``ends'' meet.

The transport of Galactic protons inside the heliosphere was initially treated by \citet{1965P&SS...13....9P}, who demonstrated that -- in the framework of statistical physics -- the random walk of the CR particles is a Markov process that can be described by a Fokker-Planck equation (e.g., see also \citealt{1965P&SS...13..115A}, \citealt{1976JGR....81.4633F}, \citealt{1993ApJ...403..760P}, and also Sections 8.2.4--8.2.4.3 of \citealt{LeroyRancoita2016}, and references therein). However, in most applications the effect of solar modulation was treated using the simplest force-field approximation~\citep{gleeson1967,1968ApJ...154.1011G} in which the diffusion tensor is approximated by a scalar and the resulting modulation effects are expressed with a spherically symmetric modulated differential number density. This was ``matched'' by the uniform Leaky-Box model for Galactic propagation -- a combination, which dominated the CR interpretation landscape in the second part of the 20th century and in the beginning of the 21st. With the development of sophisticated Galactic propagation models, the force-field approximation became the Achilles' hill of CR astrophysics. More advanced models did exist -- including those accounting for the so-called ``charge drift effect'' \imos{\citep[e.g.,][]{Jokipii77,2012ApJ...760...60B}}, whose experimental evidence was provided, for instance, by \citet{1986JGR....91.2858G} and \citet{BoellaEtAl2001}, -- but they were not as ``user friendly'' and their wide practical application was suppressed by the high ``threshold'' demand of an expertise in the heliospheric physics.



The first ever CR measurements made by Voyager 1 outside of the heliosphere, representing the figurative 0.1\% of all direct CR measurements, have an impact of many orders of magnitude larger than their ``nominal value.'' Combined with the measurements of spectra of Galactic CR species over the last two decades at various levels of solar activity, they provide a synergetic effect that enables us finally to get a grip on CR propagation in those most important $10^{-9}$ pc$^3$, which represent the heliospheric volume.

The Voyager 1 data combined with recent AMS-02, PAMELA, and earlier BESS-Polar measurements, triggered a series of papers \citep{2016PhRvD..93d3016C,2016ApJ...829....8C,2016A&A...591A..94G} aiming at the derivation of the LIS for protons and He, and producing a generalization of the modulation potential that depends on time, charge sign, and rigidity, using the force-field approximation \citep{1968ApJ...154.1011G} as a baseline. The parameterizations proposed in these papers and the correlations with the neutron monitor rate, the tilt angle of the heliospheric current sheet, and the polarity and strength of the heliospheric magnetic field they found is certainly a large step forward over the use of the simple force-field approximation in a not-so-distant past. Meanwhile, these results remain semi-phenomenological as they seek for correlations rather than solving a proper equation for particle transport in the heliosphere. Another study \citep{2016Ap&SS.361...48B} combines Voyager 1 and PAMELA data together with GALPROP model for interstellar propagation to derive the proton, He, and carbon LIS, but lacks a  proper study of the propagation parameter space for both interstellar and heliospheric propagation.    

In this paper, we use a recently developed version of a 2D Monte Carlo code for heliospheric propagation \citep[i.e., the HelMod\footnote{http://www.helmod.org/\label{helmod-link}} model,][]{Bobik2011ApJ,DellaTorre2013AdvAstro,DellaTorre2016_OneD} combined with \galprop{} to take advantage of a significant progress in CR measurements to derive the LIS for protons, helium and antiprotons. The HelMod model includes all relevant effects and, thus, a full description of the diffusion tensor. HelMod allows an accurate calculation of the heliospheric modulation for an arbitrary epoch and is fully compatible with \galprop{}.
\imos{It provides an accurate calculation of heliospheric propagation for particles with rigidities above 1 GV.}



\section{\galprop{} Model for CR Production and Propagation in the Galaxy}
\label{galprop}
The \galprop{} model for CR propagation is being continuously developed in order to provide a framework for theoretical studies of CR propagation in the Galaxy and interpretation of relevant observations (for more details see \citealt{1998ApJ...493..694M,2000ApJ...528..357M}, \citealt{1998ApJ...509..212S}, \citealt{2000ApJ...537..763S,2004ApJ...613..962S}, \citealt{2002ApJ...565..280M,2003ApJ...586.1050M}, \citealt{2006ApJ...642..902P}, \citealt{2007ARNPS..57..285S}, \citealt{2011ApJ...729..106T}, \citealt{2011CoPhC.182.1156V,2012ApJ...752...68V}, \citealt{2016ApJ...824...16J}). \galprop{} numerically solves the system of time-dependent partial differential equations describing the particle transport with a given source distribution and boundary conditions for all CR species.

In spite of its relative simplicity, the diffusion equation is remarkably successful at modeling transport processes in the ISM. The processes involved include diffusive reacceleration and convection (Galactic wind), and for nuclei, nuclear spallation, production of secondary particles and isotopes, radioactive decay, electron K-capture and stripping, in addition to energy losses due to ionization and Coulomb scattering. For CR electrons and positrons, important processes are the electron knock-on at low energies, and the energy losses due to ionization, Coulomb scattering, bremsstrahlung (with the neutral and ionized gas), inverse Compton (IC) scattering, and synchrotron emission. Secondary antiproton production in $pp$-, $pA$-, and $AA$-interactions is calculated using the results of QGSJET-IIm \citep{2015ApJ...803...54K}, a dedicated version of the QGSJET-II hadronic interaction model, while inelastically scattered protons and antiprotons are treated as ``secondary'' protons and ``tertiary'' antiprotons, correspondingly.

Galactic properties on large scales, including the diffusion coefficient, halo size, Alfv\'en velocity and/or convection velocity, as well as the mechanisms and sites of CR acceleration, can be probed by measuring isotopic abundances and spectra of primary and secondary CR species. The ratio of the halo size to the diffusion coefficient can be constrained by measuring the abundance of stable secondaries, such as, e.g., $_{5}$B. The measured abundances of radioactive isotopes ($^{10}_{4}$Be, $^{26}_{13}$Al, $^{36}_{17}$Cl, $^{54}_{25}$Mn) then allow the remaining degeneracy to be lifted resulting in the independent determination of the halo size and the diffusion coefficient \citep[e.g.,][]{1998A&A...337..859P,1998ApJ...509..212S,1998ApJ...506..335W,2001ICRC....5.1836M}. The interpretation of the peaks observed in the secondary-to-primary  ratios (e.g., $_5$B/$_6$C, [$_{21}$Sc+$_{22}$Ti+$_{23}$V]/$_{26}$Fe) around energies of a few GeV/nucleon was debated in the literature, but the reacceleration model \citep{1990acr..book.....B,1994ApJ...431..705S} with \citet{1941DoSSR..30..301K} spectrum of interstellar turbulence received a strong support from new data on the B/C ratio by PAMELA \citep{2014ApJ...791...93A} and AMS-02 \citep{PhysRevLett.117.231102}.

Closely related to CR propagation, is the production of the Galactic diffuse \gray{s} \citep{2000ApJ...537..763S,2004ApJ...613..962S} and synchrotron emission \citep{2011A&A...534A..54S,2013MNRAS.436.2127O}. Proper modeling of the diffuse \gray{} emission, including the disentanglement of the different components, is impossible without well-developed models for distributions of the interstellar gas and radiation field \citep[see, e.g.,][]{2007ARNPS..57..285S,2008ApJ...682..400P,2012ApJ...750....3A}. Global CR-related properties of the Milky Way galaxy are discussed in \citet{2010ApJ...722L..58S}.

The \galprop{} project now has nearly 20 years of development behind it. The key idea behind  \galprop{} is that all CR-related data, including direct measurements, \gray{s}, sychrotron radiation, etc., are subject to the same Galactic physics and must therefore be modeled simultaneously. The original FORTRAN90 code has been public since 1998, and a rewritten C++ version was produced in 2001. The latest major public release is v54 \citep{2011CoPhC.182.1156V}. The latest released version and supplementary datasets are available through a WebRun interface at the dedicated website. The website also contains links to all \galprop{} publications and has detailed information on CR propagation and the \galprop{} code.

In this work we use a newly developed version 55 of the \galprop{} code, which is described in \citet{PoS(ICRC2015)492}, and references therein. The current version has the possibility to vary the injection spectrum independently for each isotope. It also includes the progenitor/end-nucleus dependency tree pre-built from the nuclear reaction network and made for each species to ensure that its dependencies are propagated before the source term is generated.  This way, special cases of $\beta^-$-decay (e.g., $^{10}$Be$\to^{10}$B) are treated properly in one pass of the reaction network, instead of the two passes required before, thus providing a significant gain in speed.

\subsection{Markov Chain Monte Carlo}\label{Sect::MCMC}

Markov Chain Monte Carlo (MCMC) methods are a class of algorithms for sampling from a probability distribution, based on a Markov chain that has the desired distribution as its equilibrium distribution. The link between the target distribution of the parameters and the experimental data is given by the likelihood function. These techniques are widely applied to give posterior multi-dimensional parameter constraints from observational data and have a low computational cost, since it scales about linearly with the number of parameters. The MCMC interface to the development version of \galprop{} was adapted from CosRayMC~\citep{2012PhRvD..85d3507L} and, in general, from COSMOMC package \citep{2002PhRvD..66j3511L}, embedding \galprop{} framework into the MCMC scheme.
An iterative procedure was developed to feed the \galprop{} output into HelMod that provides modulated spectra for specific time periods to compare with AMS-02 data as observational constraints.

The basic features of CR propagation in the Galaxy are well-known, but the exact values of propagation parameters depend on the assumed propagation model and accuracy of selected CR data. 
Therefore, we use the MCMC procedure to determine the propagation parameters using the best available CR measurements.
Six main propagation parameters, that affect the overall shape of CR spectra, were left free in the scan using the 2D \galprop{} model: the Galactic halo half-width $z$, the normalization of the diffusion coefficient $D_0$ at the reference rigidity $R_D=4.5$ GV and
the index of its rigidity dependence $\delta$, the Alfv\'en velocity $V_{\rm Alf}$, the convection velocity and its gradient ($V_{\rm conv}$, $dV_{\rm conv}/dz$). It is commonly accepted that the spatial distribution of CRs only weakly depends on the chosen radial size of the Galaxy, which is set to 20 kpc. 
Besides, to correctly fit the AMS-02 proton data at low energies and to maintain a good agreement above 200 MV with Voyager 1 data (see Section \ref{Sect::LISOutsideMod}), we introduced a factor $\beta^\eta$ in the diffusion coefficient, where $\beta=v/c$ and $\eta$ was left free: the best fit value of $\eta$ is 0.91 (see Table~\ref{tbl-1}), very close to 1, so it has no effect on the LIS of the nuclei. 

Parameters of the injection spectra, such as spectral indices and the break rigidities, were also left free, but their exact values depend on the solar modulation, so the low energy parts of the spectra are tuned together with the solar modulation parameters as detailed below.
To refine the LIS description we added smoothing features for the breaks in the injection spectrum.
The numerical values of the CR source distribution parameters \citep{2011ApJ...729..106T}, $z_{\rm scale}=0.2,\alpha=1.5$, and $\beta=3.5$, remain unchanged for all scans. 


\begin{deluxetable*}{crrcrrr}
\tablecolumns{7}
\tablewidth{0mm}
\tablecaption{Propagation parameters, obtained with the MCMC posterior distributions and the GALPROP-HelMod calibration\label{tbl-1}}
\tablehead{
\colhead{N} &
\colhead{Parameter} &
\colhead{Best Value} &
\colhead{Units} &
\colhead{$1\sigma$ Mean Error} &
\colhead{\% Error} &
\colhead{Scan Range}
}
\startdata
1 & $z$ &4.0 & kpc &0.7 &18 &[1-10] \\
2 & $D_{0}/10^{28}$ &4.3 & cm$^{2}$ s$^{-1}$ &0.5 &12 &[1-10]\\
3 & $\delta$ &0.395 & \nodata &0.025  &6 &[0.3-0.9]\\
4 & $V_{\rm Alf}$ &28.6 & km s$^{-1}$ &3.0 &10 &[0-40]\\
5 & $V_{\rm conv}$ &12.4 & km s$^{-1}$ &0.8 &6 &[0-20]\\
6 & $dV_{\rm conv}/dz$ &10.2 & km s$^{-1}$ kpc$^{-1}$ &0.7 &7 &[0-20]\\
7 & $\eta$ &0.91 & \nodata & 0.05 & 5 &[0.8-1.2] 
\enddata
\end{deluxetable*}

The solar modulation is calculated using numerical functions based on HelMod (see Section \ref{Sect::PythonM}); it is implemented within the MCMC sampling procedure, after the \galprop{} run and before a comparison with the AMS-02 data is made.
At this stage, only nuclei up to $Z=14$ are included and their fragmentation and production of secondary isotopes are calculated automatically. 
Elemental abundances were derived from propagated isotopic abundances, e.g., He = $^3$He + $^4$He, with each isotope LIS independently propagated with HelMod.
Relative abundances of protons and heavier nuclei at the sources were considered in preliminary scans and revealed only a few per cent variation with respect to \galprop{} previously derived values \citep{2008ICRC....2..129M}, thus allowing us to exclude them from the main scans. Note that \citet{2016ApJ...824...16J} came to a similar conclusion.  

The high-energy break, or \textit{a change of slope}, for protons and helium highlighted by
CREAM~\citep{2010ApJ...714L..89A,2011ApJ...728..122Y}, PAMELA \citep{2011Sci...332...69A}, and AMS-02 \citep{2015PhRvL.114q1103A,2015PhRvL.115u1101A}
is computed introducing an \textit{ad hoc} spectral break in the injection spectrum.
The position of this high-energy break is tuned to be in agreement with CREAM-I data above AMS-02 range. More physical approach is to assume a change in the slope of the diffusion coefficient around 350 GV \citep{2012ApJ...752...68V}. The flattening \imos{(hardening)} of the proton and helium spectra is then reproduced if the index of the rigidity dependence of the diffusion coefficient $\delta$ is reduced above the break rigidity by $\Delta\delta\approx0.15-0.25$, dependently on the propagation model. If this is indeed the case, it can be tested once more accurate spectra for other CR species become available.

The experimental observables used in the MCMC scan include all published AMS-02 data on protons~\citep{2015PhRvL.114q1103A},
helium~\citep{2015PhRvL.115u1101A}, B/C ratio \citep{PhysRevLett.117.231102} and electrons \citep{2014PhRvL.113l1102A}, while positrons and antiprotons are excluded.
One of our current goals is to make a prediction of the antiproton spectrum \emph{based on other CR data}, while inclusion of
antiprotons \citep[AMS-02,][]{PhysRevLett.117.091103} into the scan would result in the MCMC procedure attempting to reproduce them and thus biasing our conclusions.
In the case of positrons, a significant contribution comes from sources or processes of unknown nature, therefore, their inclusion would only add free parameters that cannot be reliably constrained. The origin of the positron excess will be discussed in a follow-up paper.

\begin{figure*}[tpb!]
\centerline{
\includegraphics[width=1\textwidth]{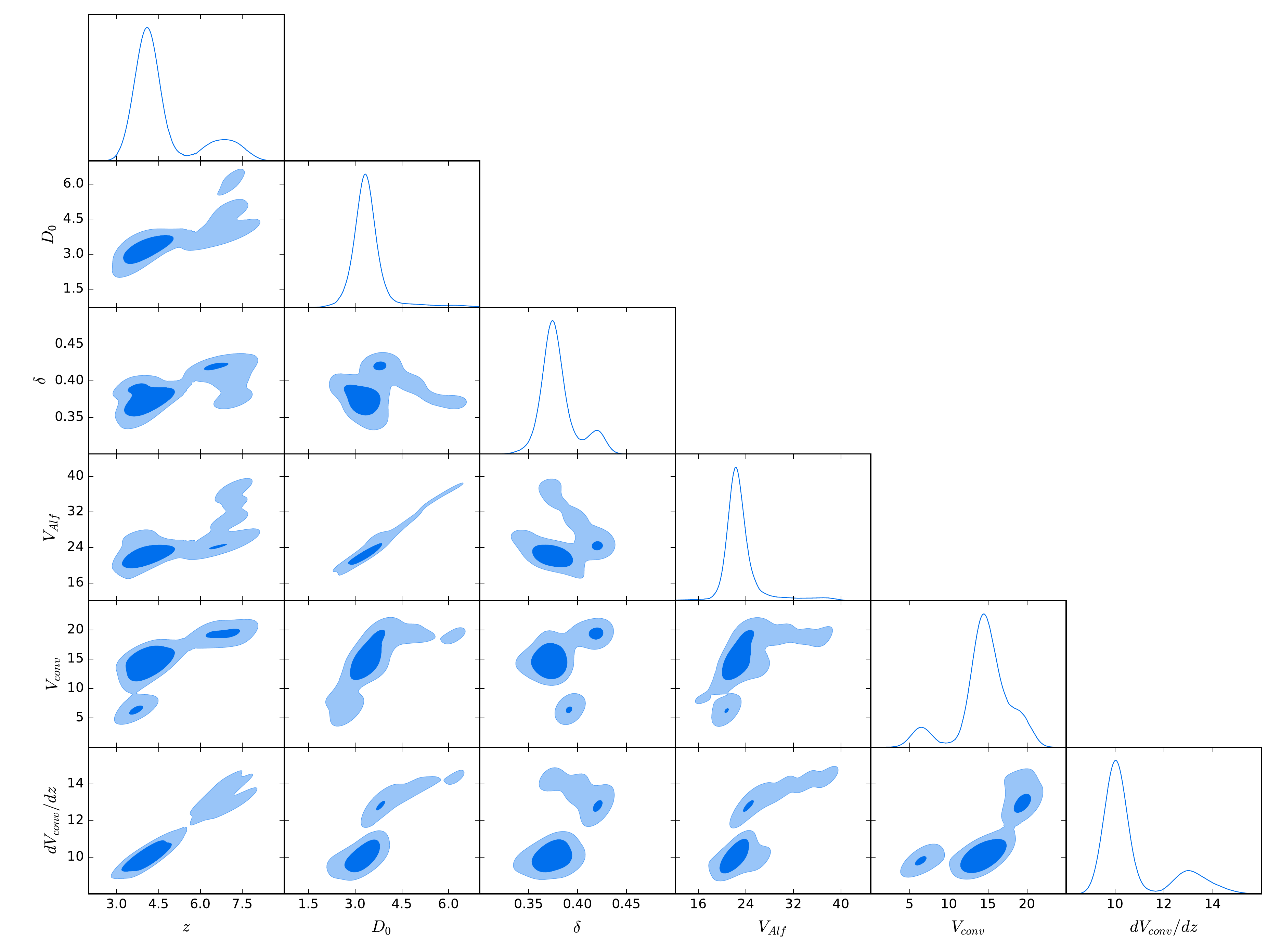}
}
\caption{MCMC matrix of two-dimensional constraints for the main propagation parameters. In off-diagonal panels, the inner contours denote $1\sigma$ (blue) and $2\sigma$ (light blue). The diagonal plots are the one dimensional probability distributions of the corresponding parameters.\label{fig:fig1}}
\end{figure*}

Simultaneous inclusion of both reacceleration and convection is needed to describe protons, particularly in the range below 20 GV where their effects on CR spectra are significant. The chosen ranges for the parameter scan, reported in Table~\ref{tbl-1}, are quite wide and allow the reacceleration and convection to be set to zero if required by the fitting algorithm. 
The goodness estimator of the parameter scan is the natural logarithm -- for computational convenience -- of the previously mentioned likelihood function,
which is built with the $\chi^2$ from all observables: hundreds of thousands of samples were generated
and the Log-Likelihood used to accept or reject each sample. The scan is terminated when a good agreement is reached.

Note that the current MCMC setup has several distinct differences from those usually employed in the literature. (i) In the current scan we use $p$, He, B/C, and $e^-$ data from AMS-02 experiment \emph{only}, i.e.\ only data $>$2 GV are used. In the case of the B/C ratio this means that we use the data above its peak rigidity. (ii) Both reacceleration and convection processes are included simultaneously. (iii) We do not use the force-field approximation. Instead, for the modulation calculations we use the HelMod routine with fixed approximate parameter values as described in detail in Section \ref{Sect::Helmod}. (iv) The MCMC procedure is used to find the best values and confidence limits for the interstellar propagation parameters and the injection spectra. The interstellar propagation was fixed after this step. (v) A grid of GALPROP models is built using small, within a few per cent, variations of the best fit parameter values found at the previous step. This model grid is used for a fine tuning of the heliospheric propagation,
more details are given in Sect.~\ref{Sect::Helmod}.

Therefore, the MCMC procedure is used only in the first step to define a consistent parameter space (Figure \ref{fig:fig1}), then a methodical calibration of the model employing the HelMod Module was performed. 
Consequently, the best values in Table~\ref{tbl-1} \emph{are not} necessarily the most probable values (MPV) obtained with the MCMC procedure, but the final values which come from the \galprop{}-HelMod combined fine tuning, that involved an exploration of the parameters space around the best values defined in the first step. The associated symmetrical mean errors are derived from the mean of a number of MCMC scans. 

\imos{The MCMC scan gives a non-zero value for $V_{\rm conv} = 12.4$ km s$^{-1}$. Strictly speaking, this may look like a discontinuity at $z=0$ because wind blows in both directions of the Galactic plane. However, in reality this wind speed is very small compared to the speed of even low-energy CR particles. Taking into account the coarse size of the spacial grid $dz=0.1$ kpc, this does not pose any problem for the solution.}

The injection spectrum parameters for each species, such as the indices $\gamma_i$ below and above the rigidity breaks $R_i$, have been moved together with solar modulation parameters within physical ranges in order to find best fit solutions for all the observables (see Section ~\ref{Sect::Helmod}). Several MCMC scans in $\gamma_i$ and $R_i$ were performed at this step. The resulting best parameters for protons and helium are shown in Table~\ref{tbl-2}. 
Due to the intercalibration between HelMod and \galprop{} parameters, the errors coming from the MCMC scans and shown in Table~\ref{tbl-2} could 
somewhat underestimate the total combined errors. 
The stability of the calculated LIS is tested through variations of five \galprop{} converging algorithm parameters (Table~\ref{tbl-3}).

\begin{deluxetable}{crrrrr}
\tablecolumns{6}
\tablewidth{0mm}
\tablecaption{Spectral parameters for protons, helium, and electrons
\label{tbl-2}}
\tablehead{
\colhead{Parameters} &
\colhead{$p$} &
\colhead{He} &
\colhead{$e^-$} &
\colhead{Mean Error} &
\colhead{Range}
}
\startdata
$R_{1}$ &7 GV &7 GV & 6 GV &1 GV &[4-10] \\
$R_{2}$ &360 GV &330 GV &100 GV &10 GV &[300-400]\\
${\gamma}_1$ &1.69 &1.71 &1.45 &0.06 &[1.5-2.1]\\
${\gamma}_2$ &2.44 &2.38 &2.75 &0.04 &[2.1-2.7]\\
${\gamma}_3$ &2.28 &2.21 &2.49 &0.05 &[2-2.4]
\enddata
\tablecomments{The estimate of the errors is to be considered qualitative, because of correlations between \galprop{} and HelMod, uncertainties in the shape of the injection spectra and lack of definitive data at few TV.}
\end{deluxetable}

\begin{deluxetable}{crr}
\tablecolumns{3}
\tablewidth{0mm}
\tablecaption{Parameters of the numerical scheme\label{tbl-3}}
\tablehead{
\colhead{N} &
\colhead{Algorithm parameters} &
\colhead{Values}
}
\startdata
1 & dz & 0.1 kpc \\
2 & Time step factor &0.75 \\
3 & Time step repetitions &30 \\
4 & $E_{\rm kin}$ factor &1.07 \\
5 & Time propagation interval & $70-10^9$ years
\enddata
\tablecomments{See \galprop{} Explanatory Supplement for details.}
\end{deluxetable}

\section{HelMod Model for heliospheric transport}
\label{Sect::Helmod}

The observed Galactic CR spectra at Earth vary with time accordingly to solar activity.
Their propagation in the heliosphere is affected by the outwards flowing solar wind (SW) with its embedded magnetic field and magnetic-field irregularities. The so-generated and transported heliospheric magnetic field is characterized by both the large scale structure (SW expansion from a rotating source) and low scale irregularities that vary with time according to the solar activity (e.g., by modification of SW velocity or local perturbations related to coronal mass ejection [CME]). The SW expansion causes CRs to propagate in a moving medium which is accounted for in the transport equation by the diffusion and adiabatic energy loss terms. In addition, according to the original formulation by \citet{parker58}, the heliospheric magnetic field (HMF) follows an Archimedean spiral that causes charged particles (i.e., CRs) to experience a combination of gradient, curvature and current sheet drifts \citep[see e.g.][]{Jokipii77},
 whose experimental evidence was provided, for instance, by~\citet{1986JGR....91.2858G} and \citet{BoellaEtAl2001}.
The overall effect of heliospheric propagation on the spectra of Galactic CRs is called solar modulation.

CR propagation in the heliosphere was first studied by \citet{1965P&SS...13....9P}, who formulated the transport equation, also referred to as Parker equation~\citep[see, e.g., discussion in][and reference therein]{Bobik2011ApJ}:
\begin{align}
\label{EQ::FPE}
 \frac{\partial U}{\partial t}= &\frac{\partial}{\partial x_i} \left( K^S_{ij}\frac{\partial \mathrm{U} }{\partial x_j}\right)\\
&+\frac{1}{3}\frac{\partial V_{ \mathrm{sw},i} }{\partial x_i} \frac{\partial }{\partial T}\left(\alpha_{\mathrm{rel} }T\mathrm{U} \right)
- \frac{\partial}{\partial x_i} [ (V_{ \mathrm{sw},i}+v_{d,i})\mathrm{U}],\nonumber
\end{align}
where $U$ is the number density of Galactic CR particles per unit of kinetic energy $T$, $t$ is time, $V_{ \mathrm{sw},i}$ is the solar wind velocity along the axis $x_i$, $K^S_{ij}$ is the symmetric part of the diffusion tensor,
$v_{d,i}$ is the particle magnetic drift velocity (related to the antisymmetric part of the diffusion tensor),
and finally $\alpha_{\mathrm{rel} }=\frac{T+2m_r c^2}{T+m_r c^2} $, with $m_r$ the particle rest mass in units of GeV/nucleon.
The processes included in Eq.~(\ref{EQ::FPE}) are extensively discussed in the literature~\citep[see, e.g.,][and reference therein]{Potgieter2016}. Over the decades of its development~\citep{Gervasi1999,1999NuPhS..78...26G,2003ESASP.535..637B,Bobik2004,Bobik2006,symposium2008,DellaTorre2009,AstraArticle2011,Bobik2011ApJ,DellaTorre2013AdvAstro,DellaTorre2016_OneD}, the HelMod model was built to include all details of the treatment of individual processes, and, therefore, provides a realistic and unique description of the solar modulation. Here we provide a short description of the HelMod model$^{\ref{helmod-link}}$ (version 3.0), more details can be found in \citet{Bobik2011ApJ,DellaTorre2013AdvAstro}.

It is widely accepted that components of $K^S$ parallel to the magnetic field are larger than its perpendicular components, and should be described using non-linear theories \citep[for a review see, e.g.,][]{shalchi2009},
while at high rigidities (i.e., $\gg$1 GV) the diffusion tensor should have a linear (or quasi-linear)
rigidity dependence~\citep[e.g., see ][]{Gloeckler1966,1968ApJ...154.1011G,jokipii1966,jokipii1971,perko1987,potgieter1994,Strauss2011}.
The transition between the non-linear and quali-linear regimes results in a ``flattening'' of rigidity dependence as observed, for instance, by~\citet{Palmer1982} and \citet{Bieberetal1994}.
In the present work, for rigidity greater than 1\,GV, we use a functional form with a rigidity dependence following the one presented in~\citet{BurgerHattingh1998}:
\begin{equation}\label{EQ::KparActual}
 K_{||}=\frac{\beta}{3} K_0\left[ \frac{P}{1\text{GV}}+g_{\rm low}\right] \left(1+\frac{r}{\text{1 AU}}\right),
\end{equation}
where  $K_0$ is the diffusion parameter, which depends on the solar activity and magnetic polarity, $\beta$ is the particle speed in units of the speed of light, $P=qc/|Z|e$ is the particle rigidity expressed in GV, $r$ is the
heliocentric distance from the Sun in AU, and, finally, $g_{\rm low}$ is a parameter, which depends on the level of solar activity and allows the description of  the flattening with rigidity below a few GV.

As discussed in Section 2.1 of \citet{Bobik2011ApJ}, 
\imos{the diffusion parameter, $K_0$, is a scaling factor for the overall modulation intensity. It defines the global behavior of the modulation of the particle flux in the heliosphere and its dependence on time reflects the variations of properties of the interplanetary medium (like the actual solar magnetic field transported by SW and its turbulence) during the different phases of solar cycles \citep[e.g., see Equation 4 in][]{ManuelFerreiraPotgieter2014}. $K_0$ is expressed in terms of the monthly Smoothed Sunspot Numbers (SSN); such a relationship was demonstrated to be adequate for the description on how the diffusion parameter depends on solar activity and polarity\footnote{The present form of $K_0$ accounts for the dip in the latitudinal distribution of GCR as observed by Ulysses, see discussion in Section~\ref{Sect::OutsideEcliptic}.} \citep[see also discussion in Section 2.3 of][]{BoschiniAdSR2017}.}
%
Therefore, the effective modulation experienced by CRs is related to the solar activity and polarity of the magnetic field.
This approximation is valid as long as disturbances coming from the Sun
(like, CME) are short and not very frequent, and
do not significantly affect the average behavior of the heliospheric medium.

During periods of high solar activity the rate of CMEs increases
leading to a more chaotic structure of magnetic field and stronger turbulence, thus the heliospheric
magnetic field cannot be properly described by a dipole configuration.
To improve the practical relationship between $K_0$ and solar activity, we use
the Neutron Monitor Counting Rate (NMCR). In the current work, we exploit the NMCR recorded by the McMurdo station and available through the Neutron Monitor Database~\citep{nmdbWeb} following the same fitting procedure used in Section 2.1 of \citet{Bobik2011ApJ}. The NMCR allows us to account for short-time and large-scale variations occurring during the high solar activity periods, and thus to re-scale the diffusion parameter accordingly. \imos{However, the usage of NMCR during the low solar activity periods does not result in an appreciable difference, thus we keep using SSN \citep{BoschiniAdSR2017} as the activity indicator for such periods.}

Furthermore, it has to be remarked that there is no commonly accepted theory describing a diffusion in strong turbulence that could be successfully applied to the heliosphere.
For the sake of simplicity, at solar maximum, we adopt a linear dependence of $K_{||}$ on rigidity,
i.e., $g_{\rm low}=0$ in Eq.~(\ref{EQ::KparActual}), that is in qualitative agreement with simulations performed for
strong turbulence conditions as shown in Figures 3.5 and 6.5 of~\citet{shalchi2009}. Due to the lack of data, the transition  periods between low and high activity are estimated
using a smooth function that is assigned $g_{\rm low}=0$ during the high activity and becomes $g_{\rm low}=0.3$ during the low activity periods. 
Such intermediate periods are discussed in Section~\ref{intermediate}.

In this model, the spatial dependence shows a radial dependence $\propto r$ and is consistent 
with the one used in~\citet{DellaTorre2013AdvAstro}, but has no latitudinal
dependence~\citep[see also discussions in][]{JokipiiKota89,McDonald1997,Strauss2011}.
The perpendicular diffusion coefficient is taken to be proportional to $K_{||}$ with a ratio
$K_{\perp,i}/K_{||}=\rho_i$ for both $r$ and $\theta$ $i$-coordinates~\citep[e.g., see ][and references therein]{potgieter2000, BurgerHattingh1998}.
At high rigidities, this description is consistent with quasi-linear theories (QLTs).
\citet{Palmer1982} constrains the value of $\rho_i$ between 0.02 and 0.08 at Earth.
We found best agreement at $\rho_i\approx 0.06$.
As remarked in~\citet{DellaTorre2013AdvAstro}, in this description $K_{||}$
has no latitudinal dependence and a radial dependence $\propto r$;
nevertheless, the reference frame transformation between the field aligned to the spherical
heliocentric frame~\citep[see, e.g.,][]{burg2008} introduces a dependence on the polar angle.
As was shown in \citet{DellaTorre2013AdvAstro}, this is sufficient to explain the
latitudinal gradient observed by Ulysses during the latitudinal \textit{fast scan}
in 1995~\citep[see e.g.][]{Heber1996,Simpson1996}.

\emph{In the present work}, we use the drift model originally developed by~\citet{Potgieter85}
and refined using definitions of Parker's magnetic field with polar correction
as reported in~\citet{DellaTorre2013AdvAstro}~\citep[see also][for a discussion about modified Parker's magnetic field]{Raath2016}:
\imos{
\begin{equation}
\vec{v}_d = f(\theta)\vec{v}_{\rm dr} + \vec{v}_{\rm HCS},
\end{equation}
where  $\theta$ is the solar colatitude, $\vec{v}_{\rm dr}$ is related to the large scale structure of HMF, $f(\theta)$  accounts for the effects of wavy neutral sheet on transport properties in the large scale structure of HMF and, finally, $\vec{v}_{\rm HCS}$ describes the drift velocity along the neutral sheet \citep[see Section 4 of ][]{Bobik2011ApJ}.}
Since during the high activity period the heliospheric magnetic field is far from
being considered regular, in this work we introduced a correction factor that
suppresses any drift velocity at solar maximum.
For the sake of completeness we should note that the presence of turbulence
in the interplanetary medium should reduce the global effect of CR drift in the heliosphere
(see, e.g., discussion in~\citealt{Minnie2007}) and this is usually
incorporated introducing a \textit{drift suppression factor}~\citep[see, e.g.,][]{Strauss2011} that is effective at rigidity below 1 GV.

We compute the CR propagation from the Termination Shock (TS)
down to \imos{the Earth orbit} using a Monte Carlo approach, i.e.,
the HelMod Monte Carlo code~\citep{Bobik2011ApJ} that solves the
two-dimensional Parker equation for CR transport through the heliosphere.
HelMod code applies the stochastic integration to a set of stochastic differential equations (SDE),
which are fully equivalent to Eq.~(\ref{EQ::FPE})~\citep[see a discussion in, e.g.,][]{Bobik2011ApJ,DellaTorre2016_OneD}.
In this scheme, quasi-particle objects evolve \emph{backward-in-time}
from the location of the detector, i.e., from Earth, back to the TS.
The modulated spectrum is then obtained by averaging the evaluated LIS fluxes, which takes into account the reconstructed rigidity at the heliospheric
boundary~\citep[see Section 4.1.2 in][]{DellaTorre2016_OneD}.

The goodness of the model is evaluated using the $\chi^2$ minimization~\citep[see, e.g., Section 15.1 of ][]{NumericalRecipes1992}:
\begin{equation}\label{eq:Helchi2}
 \chi^2 = \sum_i\frac{ \left[J_\mathrm{HelMod}(R_i)-J_\mathrm{exp}(R_i)\right]^2}{\sigma^2_{i}},
\end{equation}
where $J_\mathrm{HelMod}$ is the differential intensity evaluated using the HelMod code,
$J_\mathrm{exp}$ is the observed differential intensity, $R_i$ is the average rigidity of the $i$th rigidity bin of
the differential intensity distribution, and $\sigma_{i}$ is the experimental error corresponding to the $i$th rigidity bin.
This quantity is evaluated for rigidities below 20 GV.
We minimize $\chi^2$ for all selected experiments for both high and low levels of solar activity.

Among the HelMod parameters described before, only two of them are directly involved in the determination of a LIS: $\rho_i$ and $g_{\rm low}$. In fact,
$\rho_i$ modifies the absolute scale of modulation intensity up to high rigidities; since this parameter also influences latitudinal gradients, its value is constrained in such a way that the obtained latitudinal gradients are in agreement with those found by Ulysses and presented in Section~\ref{Sect::OutsideEcliptic}. The maximum value of $g_{\rm low}$ refines modulated differential intensity in the low rigidity range ($<3$ GV), involving mainly low activity periods.

The procedure of intercalibration between HelMod and \galprop{} is set up as follows. In the initial step, starting LIS are coming from a previous study~\citep[i.e.,][]{Bobik2011ApJ}. 
(i) Minimization of HelMod parameters in Table~\ref{tbl-Hel}, Eq.~\eqref{eq:Helchi2}, is done on a large set of proton spectra measured along solar cycles 23-24~\citep[i.e.,][]{bess_prot,AMS01_prot,PamelaProt2013,2014A&A...569A..32M,2015PhRvL.114q1103A,BESS2007_Abe_2016}.
(ii) 
The HelMod-tuned modulation obtained in the previous step is used with the \galprop{} MCMC procedure, described in Section \ref{Sect::MCMC}, that scans Galactic propagation parameters providing a new set of LIS (protons, helium nuclei, and antiprotons).
(iii) 
The new calculated LIS are provided as input for a step one minimization (i), producing new modulation functions. 

Once the loop (i)-(ii)-(iii) is completed a couple of times, step (iii) is replaced with step (iv) as described below, so that the loop now includes steps (i)-(ii)-(iv). In step (iv), a grid of \galprop{} models, obtained from variations of Table~\ref{tbl-2} parameters within a few \% with respect to the best values from MCMC, is tested.
The proton LIS from the model which shows the best overall agreement with AMS-02 data is given as input for the HelMod minimization in step one (i) again. The tuning between HelMod and \galprop{} ends when the difference between the LIS in two subsequent cycles is below 2\%.
In Table~\ref{tbl-Hel} the third column gives the 68\% confidence interval for each of the listed HelMod parameters.

In addition, we have investigated the uncertainty in the HelMod code outputs taking into account that the final results can be affected by the assumed size of the heliosphere, numerical uncertainties, and the goodness of HelMod parameters.
In the current calculation, we assume a static and spherical heliosphere with TS located at 100 AU \citep{Bobik2011ApJ}.
Note that Voyager 1, 2 observations point to a dynamic TS that is moving inward/outward in the heliosphere \citep{Stone2005,RichardsonWang2011}, while numerical models indicate that this TS movement could be as large as $\sim$20 AU over a complete solar cycle \citep[see a discussion in][and references therein]{ManuelFerreiraPotgeter2015}.
In the HelMod framework, this is equivalent to a rescaling of the real size of the heliosphere to a reference size of 100 AU. For example, if a location of the TS is changing by 10 AU \citep[see, e.g.,][]{WashimiEtAl2011}, the Earth's position in the rescaled HelMod heliosphere is moving by 0.1 AU.
Monte Carlo simulations show that such small variations of the location of Earth do not affect the solutions, thus for the rest of the paper we assume that the detector is located at 1 AU. Even though variations of the real size of the heliosphere may be important for the analysis of CR propagation near the TS, we do not consider them in this work.

The numerical uncertainties of our Monte Carlo approach were evaluated in \citet{DellaTorre2016_OneD}, who employed the Crank-Nicholson technique for the SDE integration, and found them to be less than 0.5\% at low rigidities. The large number of simulated events ensure that the statistical errors are negligible compared to the systematic uncertainties.
Finally, we evaluate the probability distributions of HelMod parameters assuming $\Delta\chi^2$ with respect to the minimizing configuration that is compatible with 68\% confidence interval \citep[see, e.g., Section 15.6 of][]{NumericalRecipes1992}. The total errors corresponding to this confidence interval are quoted in Section~\ref{Sect::results}, while the intervals obtained from the computed errors are reported in the third column of Table~\ref{tbl-Hel}. Note that the errors increase at lower rigidities.

\begin{deluxetable}{crc}
\tablecolumns{3}
\tablewidth{0mm}
\tablecaption{Parameters of the HelMod Model tuned in the LIS Scans\label{tbl-Hel}}
\tablehead{
\colhead{HelMod parameters} &
\colhead{Values} &
\colhead{Parameter range}
}
\startdata
$\rho_i$ 		& 0.06 & 0.055--0.064 \\
$g_{\rm low}$ at solar minimum	& 0.3	& 0.0--0.5
\enddata
\end{deluxetable}


\subsection{HelMod Python Module for \galprop}\label{Sect::PythonM}

\begin{figure}[tb]
\centerline{
\includegraphics[width=0.49\textwidth]{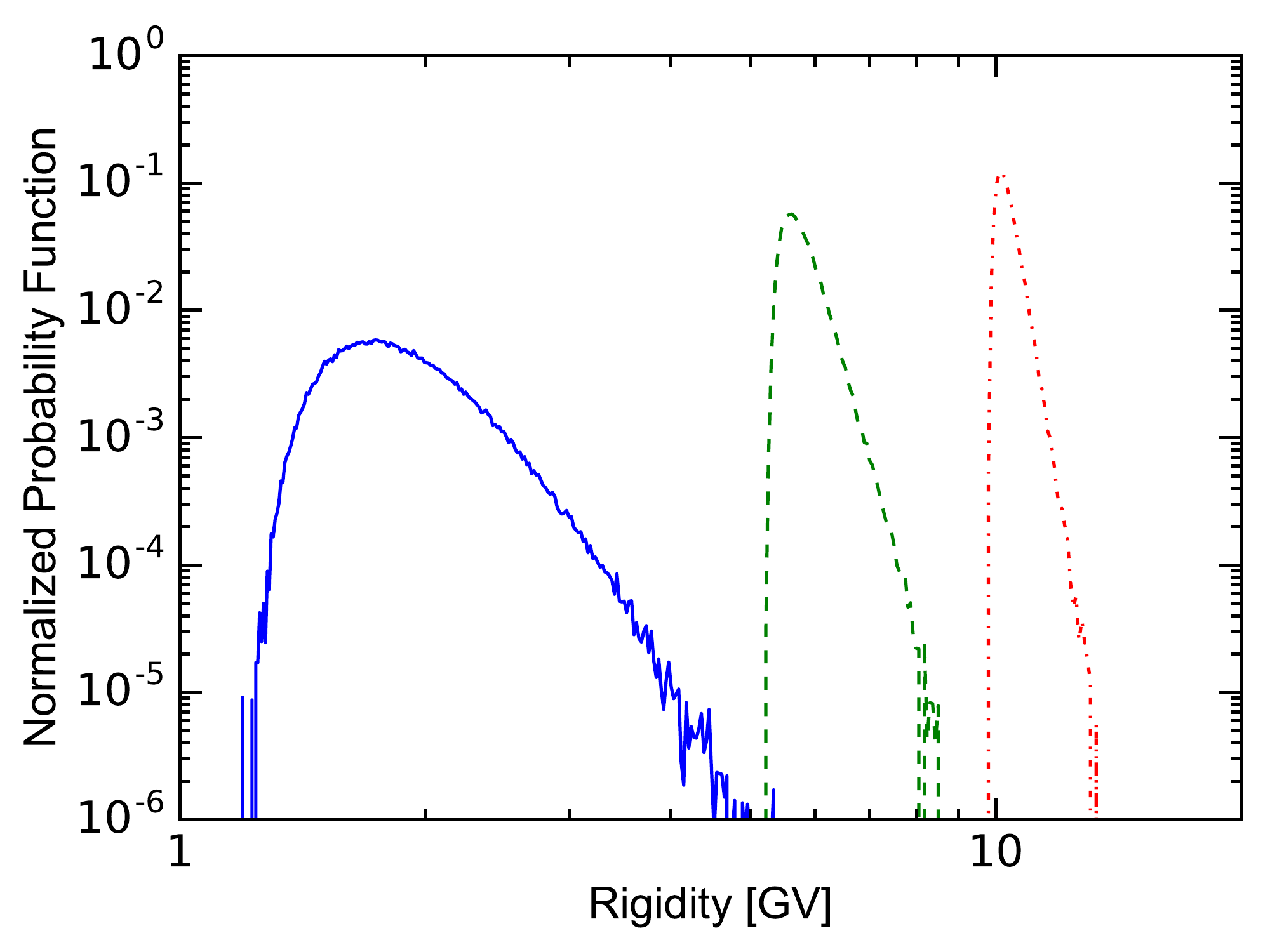}
}
\caption{The computed normalized probability function $G(R_0|R)$ for $R_0=1.1,5.1,9.7$ GV (left to right) evaluated for AMS-02 proton binning during the period 2011--2014, see text for details.
}
\label{fig:PythonModule_G}
\end{figure}

The SDE integration with HelMod results in quite an expensive effort from the computational point of view since minimization of uncertainties requires a simulation of a considerable number of events propagating from Earth to the heliospheric boundary.
The modulated spectrum is usually evaluated directly from the numerical integration using the procedure described in \citet[and references therein]{DellaTorre2016_OneD} that forces a new simulation run for each LIS to be tested.
On the other hand, the Monte Carlo integration allows us to evaluate the normalized probability
function $G(R_0|R)$ that gives a probability for a particle observed at Earth with a rigidity $R_0$ having a rigidity $R$ at the heliospheric boundary.
The modulated spectrum at specific rigidity $R_0$ is proportional to~\citep[see, e.g.,][]{PeiBurger2010}:
\begin{equation}\label{eq::PyMod_modulation}
 J_{\rm mod}(R_0)= \int_0^\infty J_{\rm LIS}(R)G(R_0|R)dR
\end{equation}
Once $G(R_0|R)$ is evaluated it is possible to obtain the modulated spectrum directly from $J_{\rm LIS}$ provided by \galprop.
For illustration, in Figure~\ref{fig:PythonModule_G} we show the computed
normalized probability function for $R_0=1.1,5.1,9.7$ GV evaluated for protons
during the period 2011-2014, equivalent to the data taking period of released AMS-02 data~\citep{2015PhRvL.114q1103A}.

To simplify the calculations,
we developed a python script that reads the \galprop{} output and provides the modulated spectrum for periods of selected experiments. The calculation of propagation in the heliosphere is substituted by the integration of Eq.~(\ref{eq::PyMod_modulation}) with the normalized probability functions, which are pre-evaluated using the HelMod code as described in the previous Section. This method dramatically accelerates the modulation calculations while provides the accuracy of the full-scale simulation.

The normalized probability functions were evaluated for several CR species ($p$, He, B/C, $e^-$) as described in Section~\ref{Sect::MCMC}. A fine tuning of \galprop{} and HelMod parameters was done using proton simulations and minimization of $\chi^2$ for a set of selected experiments running from 1997 to 2015~\citep{bess_prot,AMS01_prot,PamelaProt2013,2014A&A...569A..32M,2015PhRvL.114q1103A,BESS2007_Abe_2016}.

The HelMod python module can be downloaded from a dedicated website$^{\ref{helmod-link}}$ or used on-line. It reads the \galprop{} output format (FITS) and provides a modulated spectrum for a specified period of time. While the heliospheric propagation is fixed by using the provided functions $G(R_0|R)$, the LIS spectrum can be specified by a user. The output rigidity binning is chosen to be compatible with CR experiments in the specified period, alternatively the AMS-02 rigidity binning is chosen as the standard for results not directly associated with observational data.

\section{Interstellar propagation}

The results of the \imos{calculations} show that simultaneous inclusion of diffusion, convection, and reacceleration is required to reproduce AMS-02 measurements, while plain diffusion scenarios are excluded \citep{2007ARNPS..57..285S}. 
Since spectra of $p$, He, and heavier nuclei are well-reproduced in the standard interstellar propagation model, they can be used to put constraints on heliospheric propagation. On the other hand, having the heliospheric modulation constrained reduces the uncertainties associated with the interstellar
propagation (Tables~\ref{tbl-1}, \ref{tbl-2}) and provides up to an order of magnitude improvement in the accuracy compared to other analyses
\citep{2001ApJ...555..585M,2002astro.ph.12111M,2006ASSL..339.....D,Stanev2010,2011JCAP...03..051C,2011ApJ...729..106T,2012A&A...539A..88C,2012A&A...544A..16T, Masi2013}. For example, estimates of the index of the rigidity dependence of the diffusion coefficient $\delta$ in the literature span from $\delta=0.33-0.5$, for Kolmogorov and Kraichnan spectra of interstellar turbulence, to $\delta=0.6-0.9$ for plain diffusion models. In our analysis, the errors associated with the determination of the major propagation parameters are reduced to $\sim5\%-10\%$.  For primaries, the overall uncertainties in their spectra are so small that they are not reported in Section \ref{Sect::DataAtEarth}; some degeneracies, which affect secondary predictions, still persist and could partially account for small discrepancies with AMS-02 antiproton data.

The combined diffusion-convection-reacceleration (DCR) model has a uniform spatial diffusion coefficient ($D_{0x}=D_{0z}$) with a single power-law index ($\delta_1=\delta_2$) in the whole rigidity range.
The index $\delta$ of the rigidity dependence of the diffusion coefficient is derived from the slope of the secondary-to-primary ratio (e.g., B/C). A fit to the AMS-02 measurements of the B/C ratio \citep{PhysRevLett.117.231102} yields 0.395, which is very close to the value $\delta=0.397\pm0.007$ found from the fit to the PAMELA data \citep{2014ApJ...791...93A} and quite close to the Kolmogorov index of 1/3. 

The acceleration and diffusion processes depend on the particle rigidity and until recently the injection rigidity spectra were assumed to be the same for all nuclei. Discrepancies in the proton and He spectra were noticed in 
CREAM \citep{2010ApJ...714L..89A,2011ApJ...728..122Y}, and PAMELA data \citep{2011Sci...332...69A}, with the AMS-02 data \citep{2015PhRvL.114q1103A,2015PhRvL.115u1101A} providing an ultimate evidence for the difference $\Delta\gamma\approx 0.07-0.08$ in the spectral indices of CR protons and He. The origin of this difference is debated in the literature, but there is no consensus yet \citep{2012ApJ...752...68V}. Fitting these data requires the injection indices to be different for different nuclei species and hints that further fine-tuning may be necessary for heavier nuclei $Z\ge6$. In turn, this may affect the calculation of the B/C ratio and the propagation parameters; other effects, such as production of secondary He, N, C etc.\ from spallation of heavier nuclei are automatically taken into account in \galprop{}.
As an example, in our propagation runs, the primary helium accounts for 94\% of the total around 1 GeV/nucleon and 98\% at 100 GeV/nucleon, whereas the primary carbon accounts for 74\% of the total around 1 GeV/nucleon and 87\% at 100 GeV/nucleon. Therefore, precise measurements of heavier nuclei by AMS-02 will impose tighter constraints on CR production and propagation.

\begin{figure}
\begin{center}
 \includegraphics[width=0.49\textwidth]{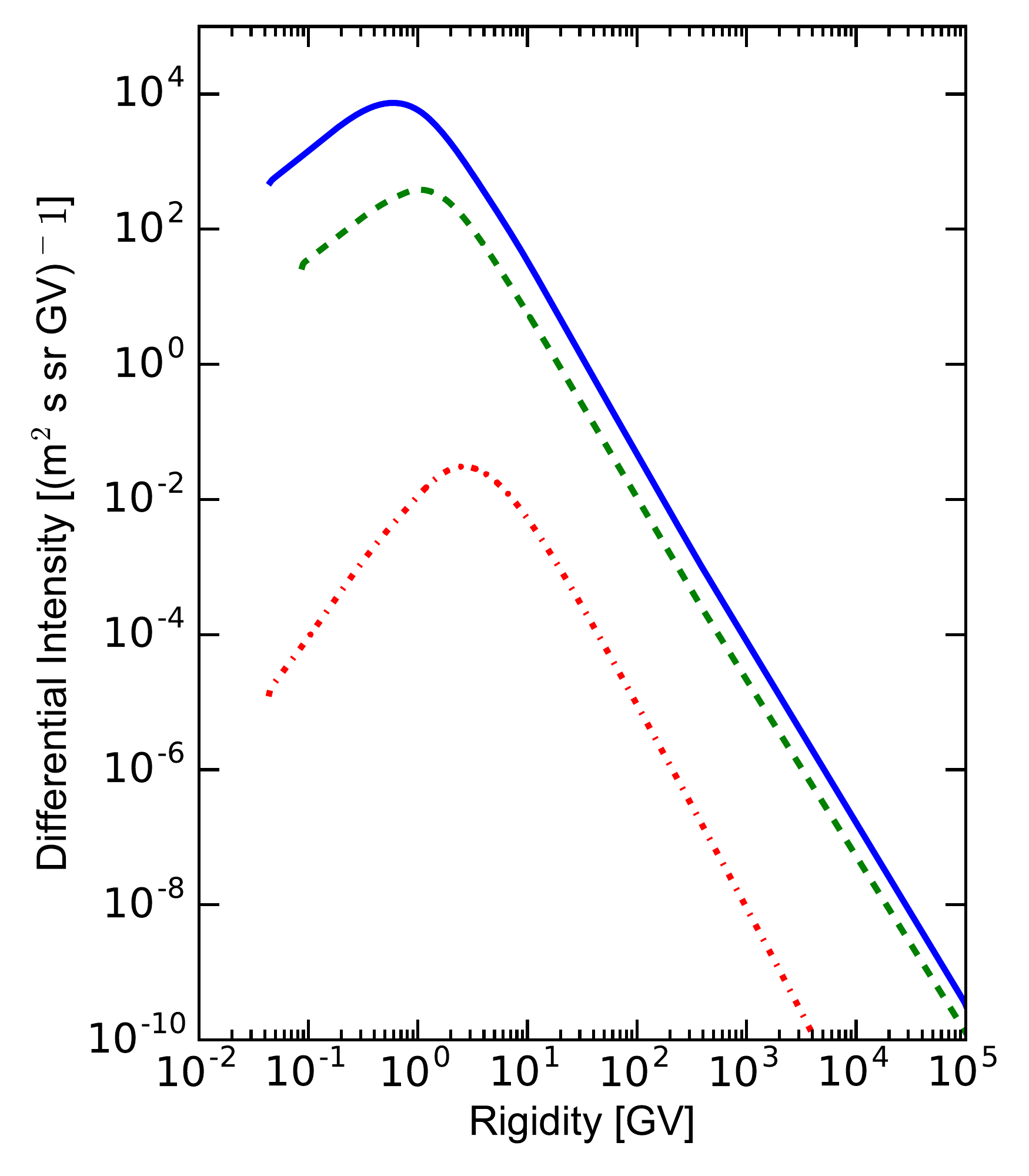}
 \caption{The local interstellar spectra of CR protons (blue solid line), helium (green dashed line), and antiprotons (red dash-dot line) as derived from the MCMC procedure (see text).}
 \label{fig:GALPROPLISS}
 \end{center}
\end{figure}

\begin{figure*}[!tb]
\centerline{
\includegraphics[width=0.49\textwidth]{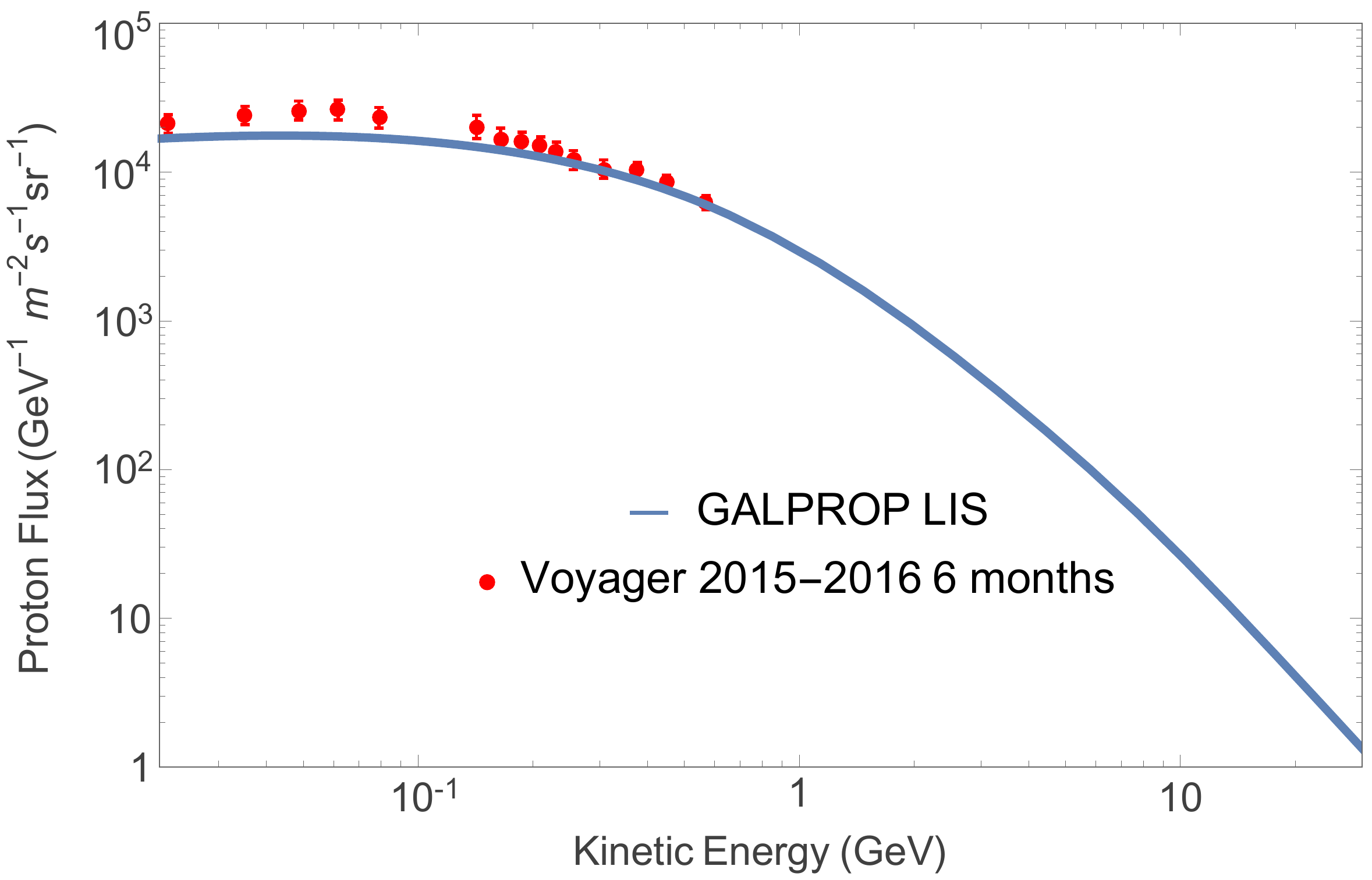}\hfill
\includegraphics[width=0.49\textwidth]{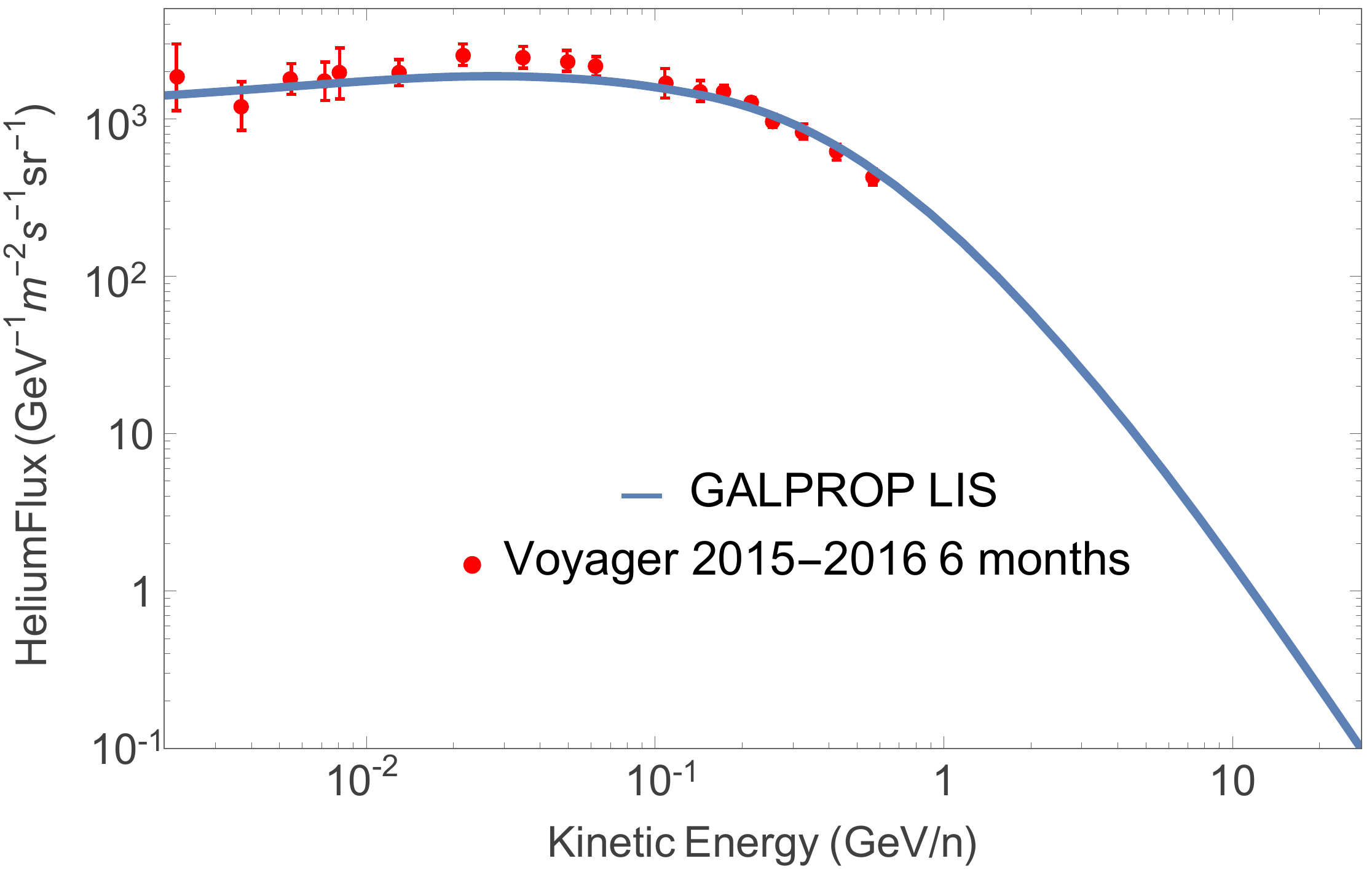}
}
\caption{The DCR best set proton LIS (left) and He LIS (right) shown with blue curves are compared with Voyager 1 2015-2016 monthly averaged data.}
\label{fig:Voyager_P_He}
\end{figure*}

The $p$, He, and $\bar p$ LIS derived with the described MCMC procedure and HelMod-\galprop{} intercalibration (using all nuclei up to $Z=28$), are shown in Figure~\ref{fig:GALPROPLISS} as functions of rigidity and tabulated in the Supplementary Material in the Appendix, Tables~\ref{Tbl-ProtonLIS}-\ref{Tbl-AntiprotonLIS}. A comparison with the data is discussed in detail in Sections~\ref{Sect::results} and \ref{Sect::LISOutsideMod}.
In addition to the tabulated data, we provide analytical fits to the derived LIS. The fit to the antiproton LIS provides an accuracy of 2--3\% for 3 GV $<R<1000$ GV and 10\% for 1.5 GV $<R<3$ GV:
\begin{align}
&F(R)\times R^{2.7} = 1.34 -4.99\times 10^{-4}\,R  \\
 &+ \frac{44.3}{12.6+R}-\frac{168}{31.1+R^{2}} +\frac{13600}{22200+R^{2}}, \quad R>1\ {\rm GV}, \nonumber
\end{align}
while the average accuracy of AMS-02 $\bar p$ data is about 10\%--20\%.
The expressions for proton and helium LIS are given in Section \ref{Sect::LISOutsideMod}.

The MCMC procedure prefers a medium size halo of 4 kpc, also favored in the past studies
\citep{2001AdSpR..27..717S}, contrary to very large or extremely small halos proposed by lepton emission models \citep{2000ApJ...537..763S,2009arXiv0910.2027G} and by synchrotron emission studies \citep{2013MNRAS.436.2127O}. The electron LIS is the subject of the followup paper dedicated to leptons and nuclei $Z>2$, although the final DCR model presented in this paper is already in a good agreement with AMS-02 electrons data. It is worth pointing out that electrons may also have a component of the same origin as the excess positrons in CRs \citep{2014PhRvL.113l1101A,2014PhRvL.113l1102A}. It may be less pronounced than in the case of positrons due to the much larger flux of electrons from conventional sources, but still deserves a more careful study.

As pointed out in \citet{2016ApJ...824...16J}, there could be a significant difference between the propagation parameters derived from the light isotopes ($p$, $\bar p$, He), and nuclei (boron to silicon). Our study does not show any evident discrepancies between the light isotopes ($p$, $\bar p$, He), and the B/C ratio (nuclei). This may be explained by the differences in the setups outlined in Section \ref{Sect::MCMC} and, in particular, by a more realistic description of heliospheric propagation used in the present analysis.

\section{Proton, Helium and Antiproton LIS}
\label{Sect::results}

As described in the previous sections, our approach combines two state-of-the-art codes, \galprop{} for interstellar propagation and HelMod for heliospheric propagation, within a single framework for the first time. AMS-02 data, which is guiding the refinement of the propagation scheme, is a vital ingredient of this approach. The converse is also true, the refined propagation scheme is beneficial for interpretation of the AMS-02 data. Combination of AMS-02 high precision data (e.g., $\sim$1\% errors for the proton spectrum) with the data taken by earlier missions (AMS-01, BESS, PAMELA) at different epochs allows the framework to be extended to account for different polarities of the solar magnetic field and for periods of high and low solar activity in cycles 23 and 24 (see Section \ref{Sect::DataAtEarth}), while at the same time providing an accurate description of the Voyager 1 spectra taken beyond the TS.

\subsection{Proton and Helium LIS outside Modulated Energy Region}\label{Sect::LISOutsideMod}

The direct measurements of proton and He LIS are now available at both low and high energies. High energy CR fluxes above 100 GV are not affected by the heliospheric modulation and their measurements provide a direct probe of the LIS. At low energies measurements of CR fluxes are provided by Voyager 1 that crossed the TS in the second half of 2012 \citep{2013Sci...341..150S,2016ApJ...831...18C}. To avoid a possible influence of turbulence that may still be present beyond the TS, we take the latest Voyager 1 2015-2016 data averaged over monthly intervals\footnote{http://voyager.gsfc.nasa.gov/heliopause/vim/monthly/index.html}.

The averages of six months of Voyager 1 data for protons and He are shown in Figure~\ref{fig:Voyager_P_He}. The error associated with each data point is chosen conservatively to be equal to the variation of the monthly average, but not smaller than 15\% of the flux value. The combined model provides a good description of proton and He LIS at low energies. The agreement with He is particularly good, while protons are slightly underestimated in the energy range between 30 and 100 MeV. Even though \galprop{} has a good description of relevant physics processes down to keV energies, we did not tune to the data below $\sim$200 MeV/nucleon (available from ACE/CRIS). We also emphasize that Voyager 1 data were not included into the MCMC scan, and a remarkable agreement between the model predictions and the LIS data is very supportive to our approach. The low energy LIS by Voyager 1 that are reproduced by \galprop{} are linked to the modulated AMS-02 data using the HelMod code.

\begin{figure*}[tb!]
\centerline{
\includegraphics[width=0.49\textwidth,height=0.26\textheight]{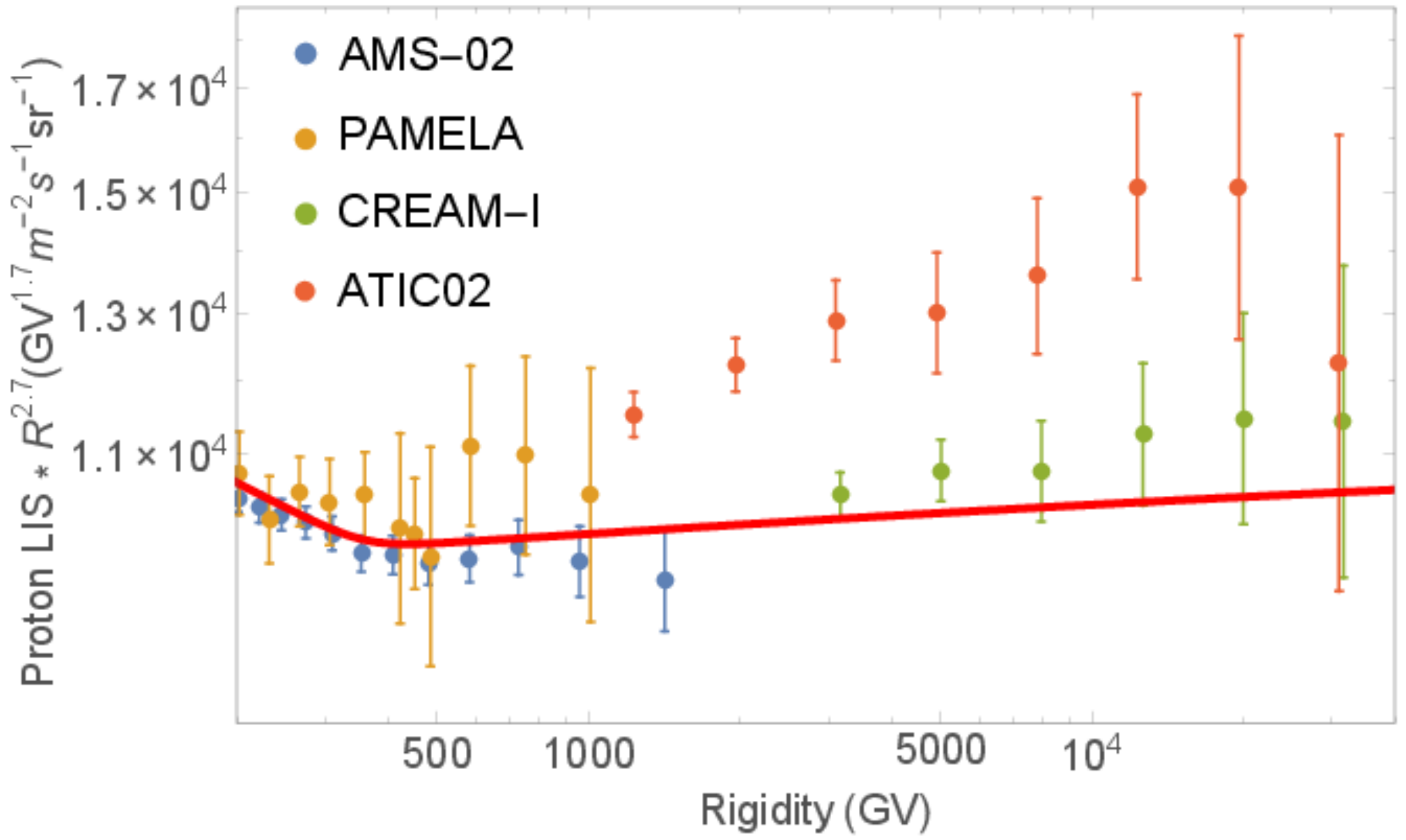}\hfill
\includegraphics[width=0.49\textwidth,height=0.26\textheight]{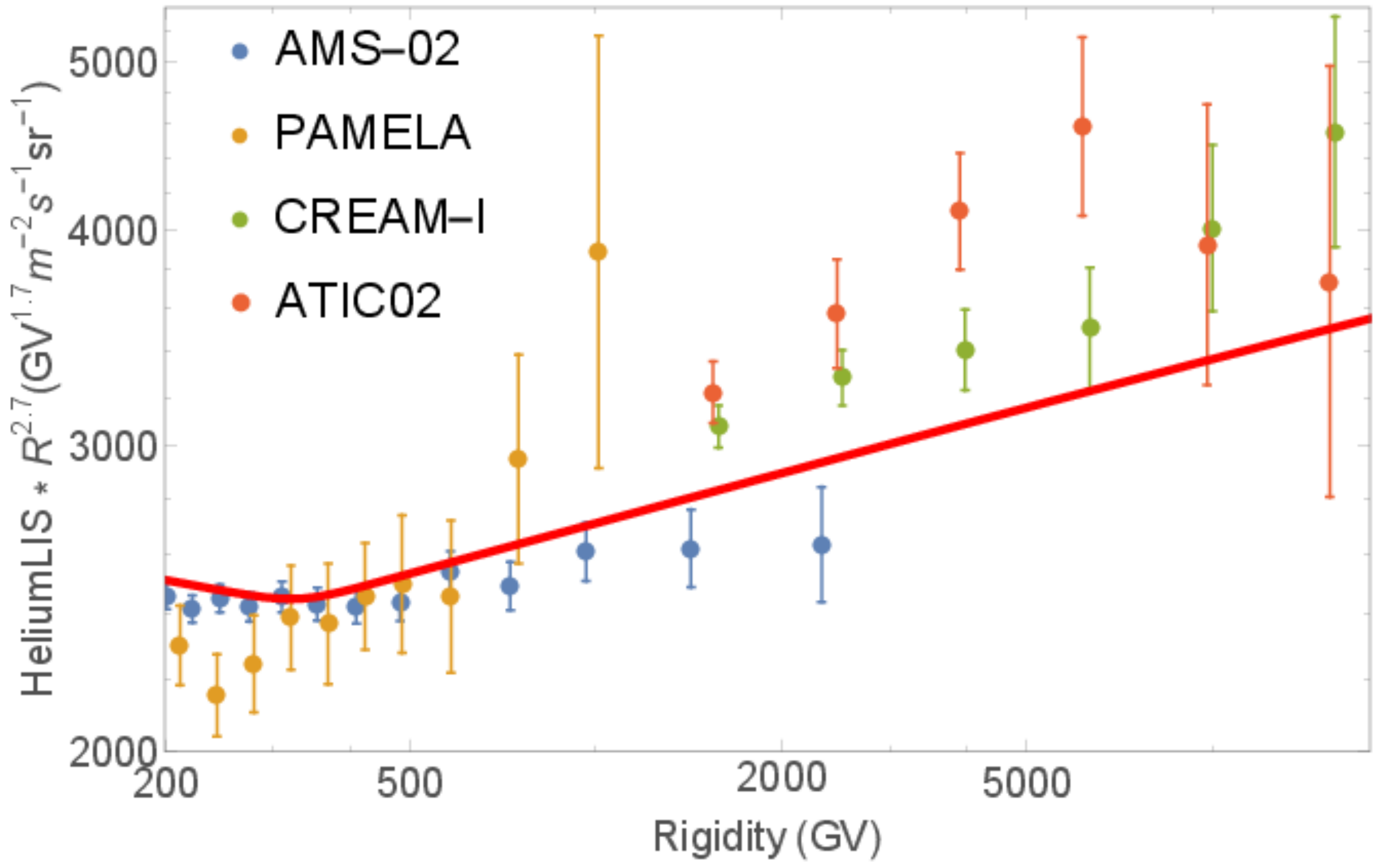}
}
\caption{The best fit proton LIS (left) and He LIS (right) are compared with high energy data by AMS-02, CREAM-I, ATIC-02, and PAMELA.}
\label{fig:high_p_He_lis}
\end{figure*}

At high energies, where CR fluxes are not affected by the heliospheric modulation, we use AMS-02 data up to $\sim$2 TV and extend the rigidity range to 20--30 TV using data taken by CREAM-I and ATIC-02 (Figure~\ref{fig:high_p_He_lis}). In this energy range the data are scarce and there is an obvious systematic discrepancy between CREAM-I and ATIC-02. Extrapolations of proton and He spectra by AMS-02, even not perfect, seem to prefer CREAM-I data which we use hereafter. It is clear from the figures that using AMS-02 data alone would lead to steeper values for the injection indices $\gamma_{3}$ at high energies. The relatively smaller error bars of CREAM-I vs.\ AMS-02 data at high energies drive the fit to result in flatter index values. The derived $\gamma_{3}$ values for protons and He (Table~\ref{tbl-2}) reflect the results of the fit to combined AMS-02 and CREAM-I data.

In addition to the plots and the tabulated data, we provide analytical functional dependence of the derived LIS (Figure~\ref{fig:GALPROPLISS}) as a function of rigidity. To provide the required accuracy (1--2\% deviations), especially in the AMS-02 range, the fit was split into two rigidity intervals, roughly below and above 1 GV. The search of the analytic solutions -- as already mentioned in Section~\ref{Sect::MCMC} -- was guided by an advanced MCMC fitting procedure such as Eureqa\footnote{http://www.nutonian.com/products/eureqa/}. The combined LIS formula looks like:
\begin{align}\label{EQ::an}
F(R)&\times R^{2.7} = \\
&\left\{
\begin{array}{ll}
\sum_{i=0}^5 a_i R^{i}, &R\le1\ {\rm GV},\nonumber\smallskip\\
b + \frac{c}{R} + \frac{d_1}{d_2+R} + \frac{e_1}{e_2+R} + \frac{f_1}{f_2+R} + g R, & R>1 \ {\rm GV},\nonumber
\end{array}
\right.\nonumber
\end{align}
where $a_i, b, c, d_i, e_i, f_i, g$ are the numerical coefficients summarized in Table~\ref{tbl-4}.

The accuracy of the low-energy expression in Eq.~(\ref{EQ::an}) is 2\% in the range $0.2\ {\rm GV}$$<$$R$$<$$1$ GV for the proton LIS. The high-energy part reproduces the numerical proton LIS calculated with \galprop{} with an accuracy of $\sim$9\% for $0.45\ {\rm GV}$$<$$R$$<$$1$ GV (i.e., $E_{\rm kin}>0.11$ GeV), where the constraints from Voyager 1 are wider than 10\%, and of $2\%$ for $R$$>$$1$ GV. The discrepancies with respect to AMS-02 data at higher energies expressed in standard deviations are virtually zero, with $0.5\sigma$ around 1--2 GV at most. In the case of Helium, the low-energy expression in Eq.~(\ref{EQ::an}) is valid in the range of $0.15\ {\rm GV}$$<$$R$$<$$2$ GV, i.e., approximately between 3 MeV/nucleon and 450 MeV/nucleon. At higher rigidities $1.5\ {\rm GV}$$<$$R$$<$$2\times 10^{4}$ GV (i.e., $>$0.3 GeV/nucleon), it reproduces the He LIS calculated with \galprop{} with an accuracy of 2\%.

The derived expressions are virtually identical, to $<$1--2\%, to numerical solutions in over 5 orders of magnitude energy interval including the spectral flattening at high energies, and are based on Voyager 1, AMS-02, and CREAM-I data.

\begin{deluxetable*}{crrrrrrrrrrrrrrr}
\tablecolumns{16}
\tablewidth{0pt}
\tabletypesize{\scriptsize}
\setlength{\tabcolsep}{1pt}
\tablecaption{Parameters of the analytical fits to the proton and He LIS\label{tbl-4}}
\tablehead{
\colhead{} &
\colhead{$a_0$} &
\colhead{$a_1$} &
\colhead{$a_2$} &
\colhead{$a_3$} &
\colhead{$a_4$} &
\colhead{$a_5$} &
\colhead{$b$} &
\colhead{$c$} &
\colhead{$d_1$} &
\colhead{$d_2$} &
\colhead{$e_1$} &
\colhead{$e_2$} &
\colhead{$f_1$} &
\colhead{$f_2$} &
\colhead{$g$}
}
\startdata
p &
$94.1$ & 
$-831$ & 
0 &
$16700$ & 
$-10200$ & 
0 &
$10800$ & 
$8590$ & 
$-4230000$ & 
$3190$ & 
$274000$ & 
$17.4$ & 
$-39400$ & 
$0.464$ & 
0
\\

He &
$1.14$ &
0 &
$-118$ & 
$578$ & 
0 &
$-87$ & 
$3120$ & 
$-5530$ & 
$3370$ & 
$1.29$ &
$134000$ & 
$88.5$ & 
$-1170000$ & 
$861$ & 
$0.03$  
\enddata
\end{deluxetable*}

\subsection{Outside the Ecliptic Plane}\label{Sect::OutsideEcliptic}

A reliable model for heliospheric modulation requires a proper modeling of CR distribution in the whole heliospheric volume including space outside the ecliptic plane and at large distances from the Sun.

Since 1990s and until 2009, the Ulysses spacecraft~\citep[see e.g.][]{Sandersonetal1995,Marsden2001,BaloghetAl2001} explored
the heliosphere outside the ecliptic plane up to $\pm 80\degree$ in solar latitude and at distances $\sim$1--5 AU from the Sun. In particular, observations of particle flux were performed using the Cosmic Ray and Solar Particle Investigation Kiel Electron Telescope (COSPIN/KET) and High Energy Telescope (COSPIN/HET). Measurements of particle fluxes with the IMP-8 (before 2006) and PAMELA (after 2006) satellites as a baseline close to Earth allow the unique derivations of spatial gradients in the inner heliosphere outside the ecliptic plane.

Ulysses observations pointed to a positive latitudinal gradient in the proton intensity (see Figure 5 in \citealp{Heber1996} and Figure 2 in \citealp{Simpson1996a}). These observations of the proton flux taken during the \textit{latitudinal fast scan} from September 1994 to August 1995, have shown (a) a nearly symmetric latitudinal gradient with the minimum near ecliptic plane, (b) a southward shift of the minimum, and (c) the intensity in the North polar region at $80\degree$ exceeding the South pole intensity. \citet{Simpson1996a} estimated a latitudinal gradient at $\sim$$0.3\%/\degree$ for protons with kinetic energy $>$0.1 GeV, while \citet{Heber1996} extended the analysis to higher energies estimating a gradient to be $\sim$$0.22\%/\degree$ for protons with kinetic energy $>$2 GeV. The minimum in the charged particle intensity separating the two heliospheric hemispheres occurs at $\sim$$10\degree$ South of the heliographic equator \citep{Simpson1996a}. An independent analysis that takes into account the latitudinal motion of Earth and IMP-8 confirms a significant ($\sim 8\degree \pm 2\degree$) southward offset of the intensity minimum for $E_{\rm kin}> 100$ MeV protons \citep{Simpson1996a}, while \citet{Heber1996} estimated a southward offset of $\approx7\degree$ independent of particle energy $<$2 GeV in their analysis. Finally \citet{Simpson1996a} estimated that the intensity in the North polar region at $80\degree$ exceeds the South pole intensity by $\sim$$6\%$ for protons of $E_{\rm kin}> 100$ MeV. However, the same study of electrons \citep{FerrandoEtAl1996} shows that the electron intensity does not show any sign of latitudinal dependence, at least, up to 2.5~GV.

Almost 11-years later, a new \textit{latitudinal fast scan} during the opposite magnetic field polarity was used to study the radial and latitudinal gradients of CR protons during the unusual solar minimum between solar cycles 23 and 24. The derived radial gradients of $2.8\pm 0.2\%/$AU for 1.9 GV protons \citep{deSimone2011,GieselerHeber2016} were similar to those found in previous studies \citep{CummingsetAl1987,McKibben1975,Heber1996}. The measured latitudinal gradient was found to be slightly negative, $-0.06\pm 0.01\%/\degree$ for 1.9 GV protons \citep[see Table A.1 in][]{GieselerHeber2016}.

\begin{figure}[tb!]
\centerline{
 \includegraphics[width=0.49\textwidth]{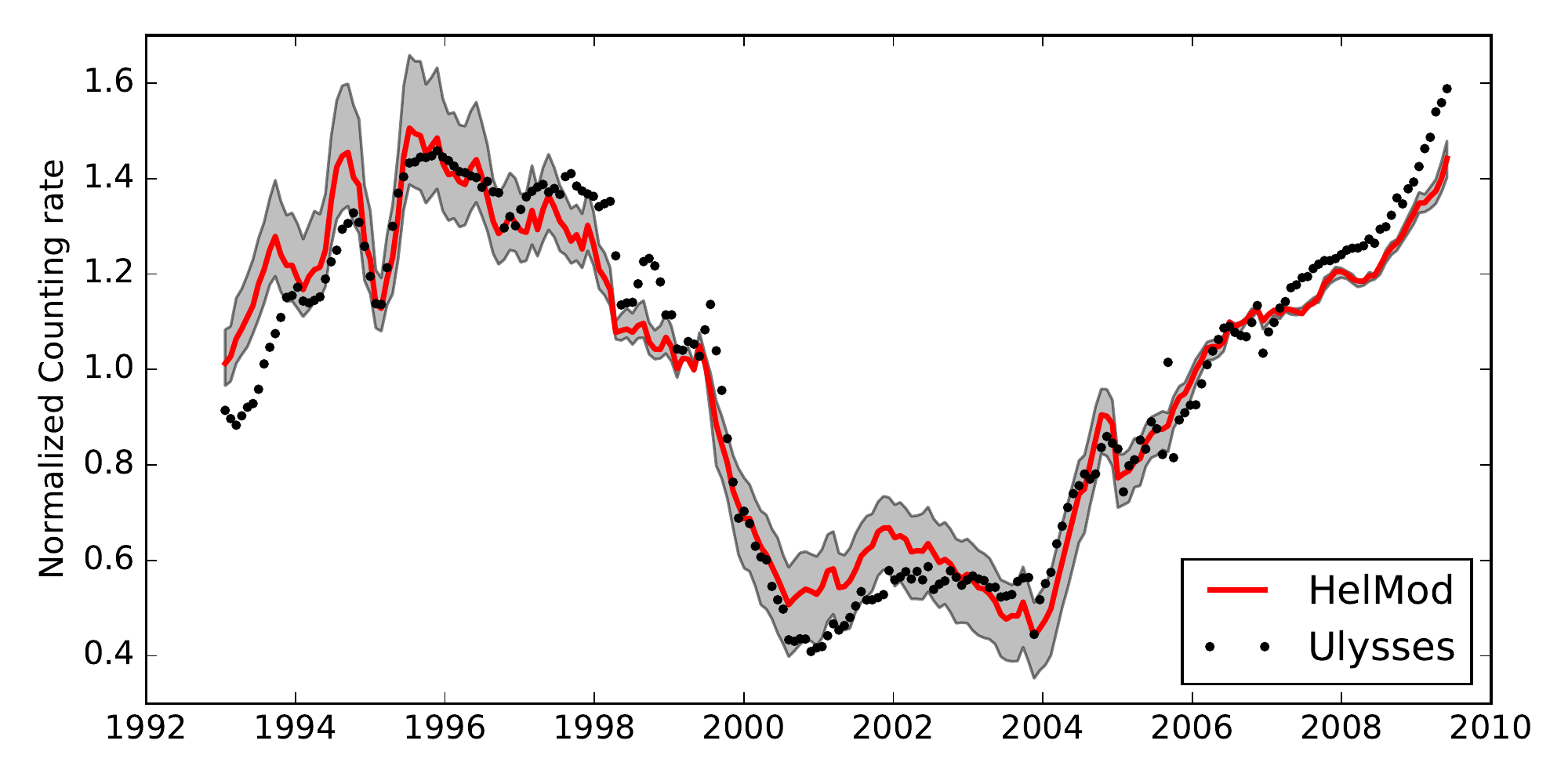}
 }
 \caption{Ulysses counting rate normalized to the average value for the KET
 coincidence channel K12 (proton in energy range 0.25-2.2 GeV/nucleon) as a function
 of time in units of Carrington rotation. Each point was averaged over a Carrington rotation.
 Solid line is the HelMod result provided for proton at 2.2 GeV for each Carrington rotation
 at same distance and solar latitude of the Ulysses spacecraft.}
 \label{fig:OutEcliptic}
\end{figure}

In \citet{DellaTorre2013AdvAstro,ICRC13_DellaTorre} we have shown that a combination of a polar modification in the description of the heliospheric magnetic field with a diffusion tensor that is independent on the solar latitude (see Section~\ref{Sect::Helmod}) is able to reproduce the measured latitudinal gradients during the low solar activity periods. In Figure~\ref{fig:OutEcliptic} we compare the Ulysses counting rate normalized to the average value with the normalized proton flux at approximately the same rigidity. Data for Ulysses were taken from Ulysses Final Archive\footnote{http://ufa.esac.esa.int/ufa}. We analyzed the data for the KET coincidence channel K12 (proton energies of 0.25--2.2 GeV/nucleon) using the Carrington Rotation average. HelMod results were provided for protons of 2.2 GeV for each Carrington Rotation at the same distance and solar latitude as the Ulysses spacecraft. The error band was evaluated using the procedure described in Section \ref{Sect::Helmod}.

\imos{In Figure~\ref{fig:OutEcliptic} one can see that the Ulysses data are qualitatively reproduced by the present model.
Both experimental data and simulations are normalized to their corresponding mean values to allow a relative comparison along the solar cycle.
The model reproduces the general features of latitudinal gradients observed during the fast scans of 1994-1995 and 2007.} Moreover, the agreement is still acceptable along the whole orbit that extends as far as $\sim$3 AU. 

\imos{We note that, the aim of Figure~\ref{fig:OutEcliptic} is only to show the qualitative agreement found between the HelMod spectrum and observation data; in fact, HelMod calculations were performed for a mono-energetic bin, while KET observations are integrated over a large energy interval \citep[e.g., see the discussion in ][]{deSimone2011}. A more quantitative comparison with Ulysses data needs to combine together simulations for several energy bins and to weight them with Ulysses response function.}

The proposed model properly accounts for latitudinal and radial gradients in the inner heliosphere. The amplitude of model uncertainties is mainly related to the ratio between the perpendicular and parallel diffusion coefficient ($\rho_i$). Larger value of $\rho_i$ leads to a more isotropic propagation, and thus to a smaller latitudinal gradient. Therefore, Ulysses data allows the value of $\rho_i$ to be reasonably constrained \citep{DellaTorre2013AdvAstro}. During the period of negative solar magnetic field polarity, this effect vanishes for positively charged particles due to the effective isotropization of particle trajectories by the global drift effects.

\subsection{Data at Earth}\label{Sect::DataAtEarth}

This Section discusses an application of the HelMod model to proton, helium, and antiproton spectra. A comparison is made with observational data for conditions of low (i.e., 1997--1998, 2006--2010) and high solar activity (i.e. 2000--2002, 2011--2013), and then with the moderate activity period, thus providing an unique model that is valid for the entire solar cycle. In this Section we show only illustrative results, more details could be found in the Supplementary Material in the Appendix.

\begin{figure*}[!p]
\centerline{
 \includegraphics[width=0.5\textwidth]{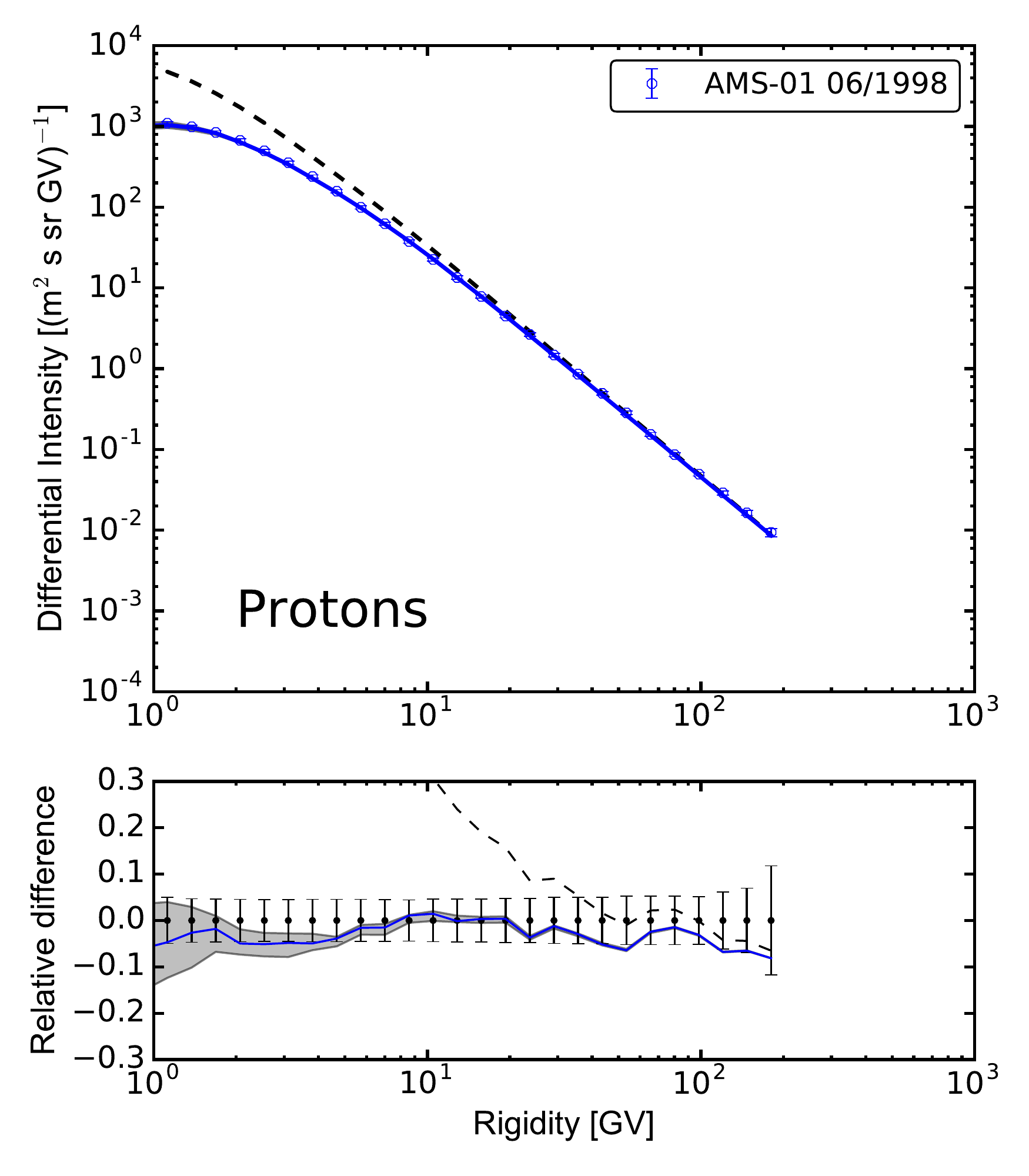}\hfill
 \includegraphics[width=0.5\textwidth]{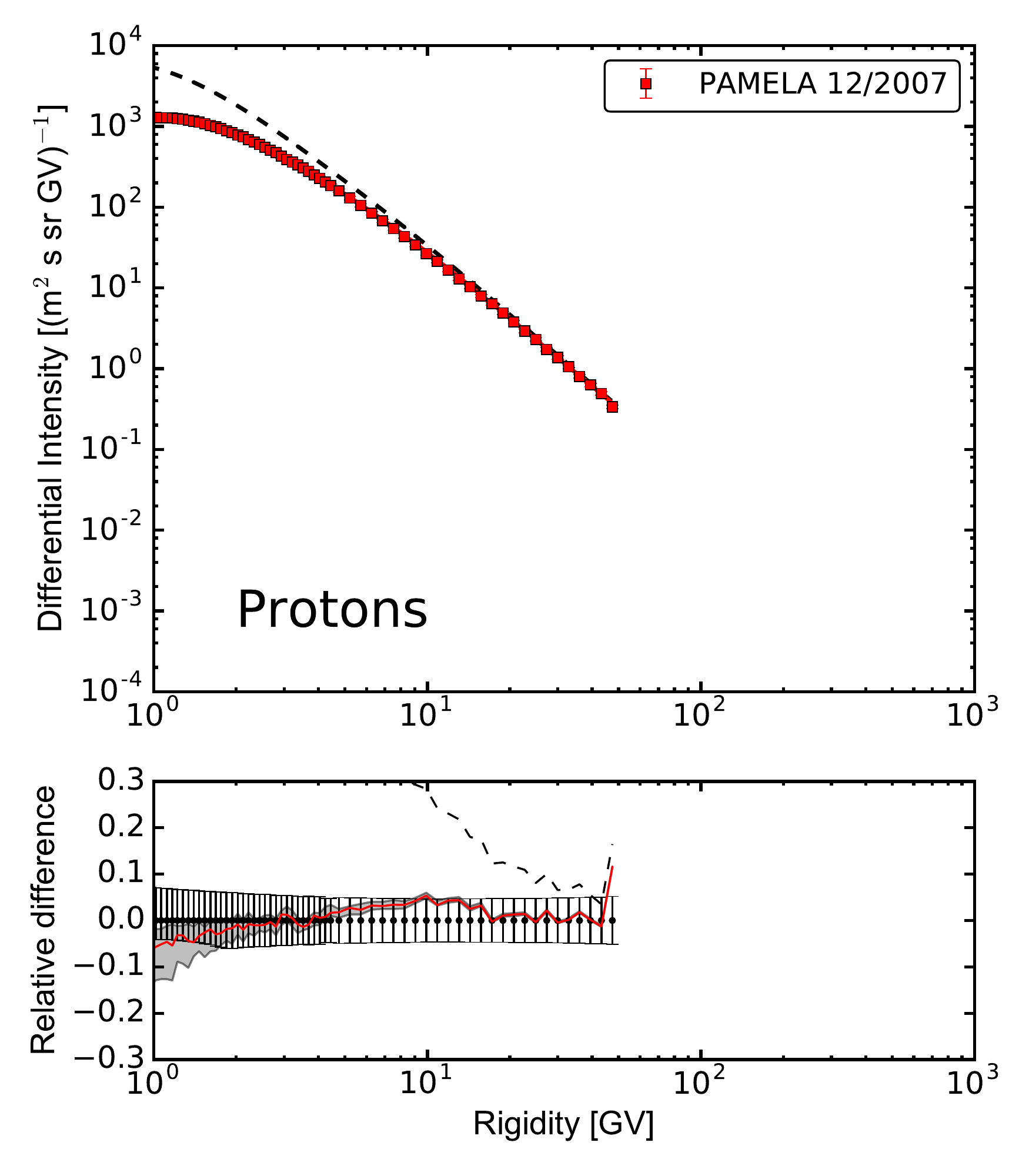}
 }
 \caption{Proton differential intensities for June 1998 (AMS-01, left) and December 2007 (PAMELA, right). Points represent experimental data, dashed line is the \galprop{} LIS, and solid line is the computed modulated spectrum. The bottom panel is the relative difference between the numerical solution and experimental data.
}
 \label{fig:Proton_Low}
\end{figure*}


\begin{figure*}[!tb]
\centerline{
 \includegraphics[width=0.5\textwidth]{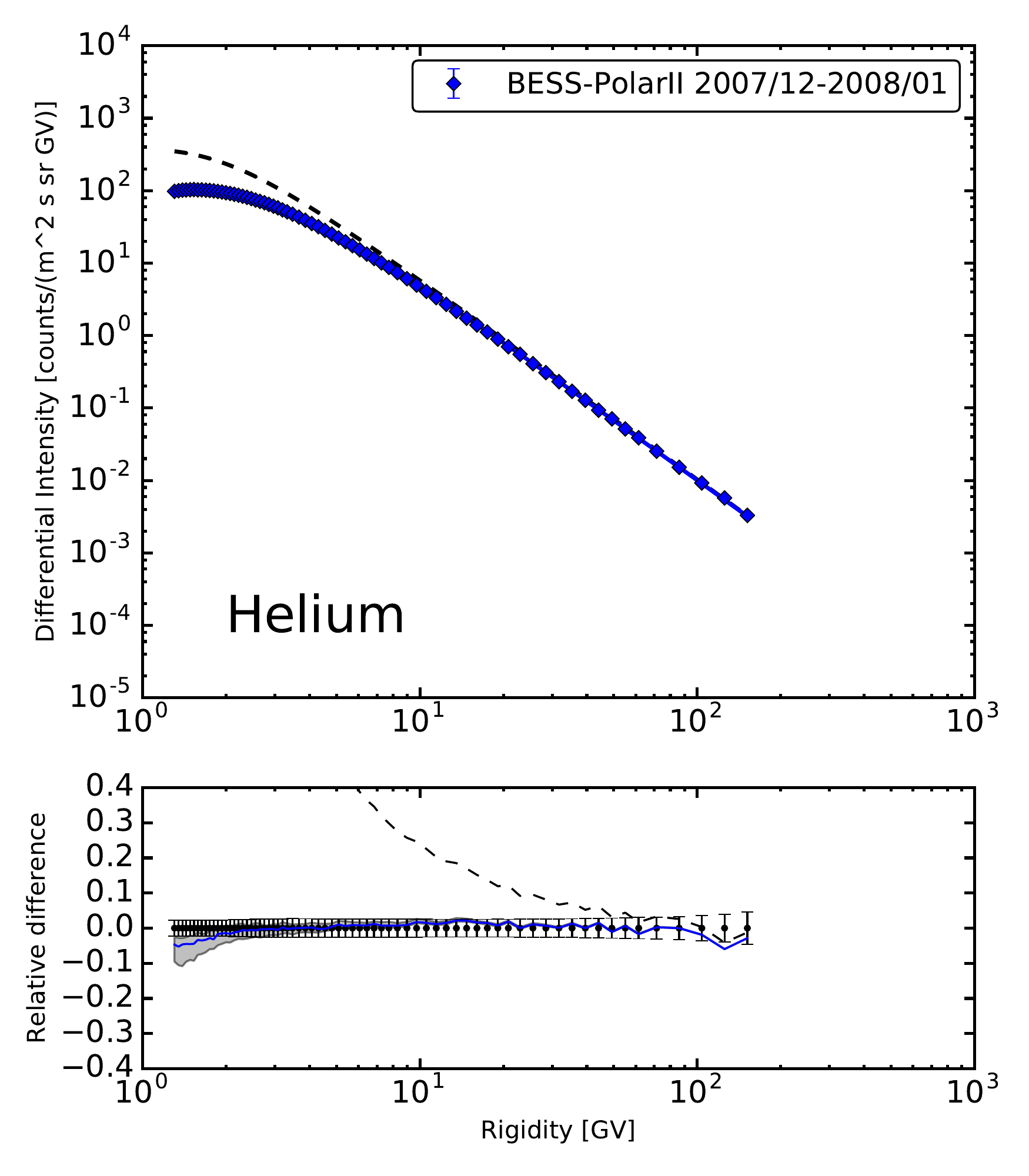}\hfill
 \includegraphics[width=0.5\textwidth]{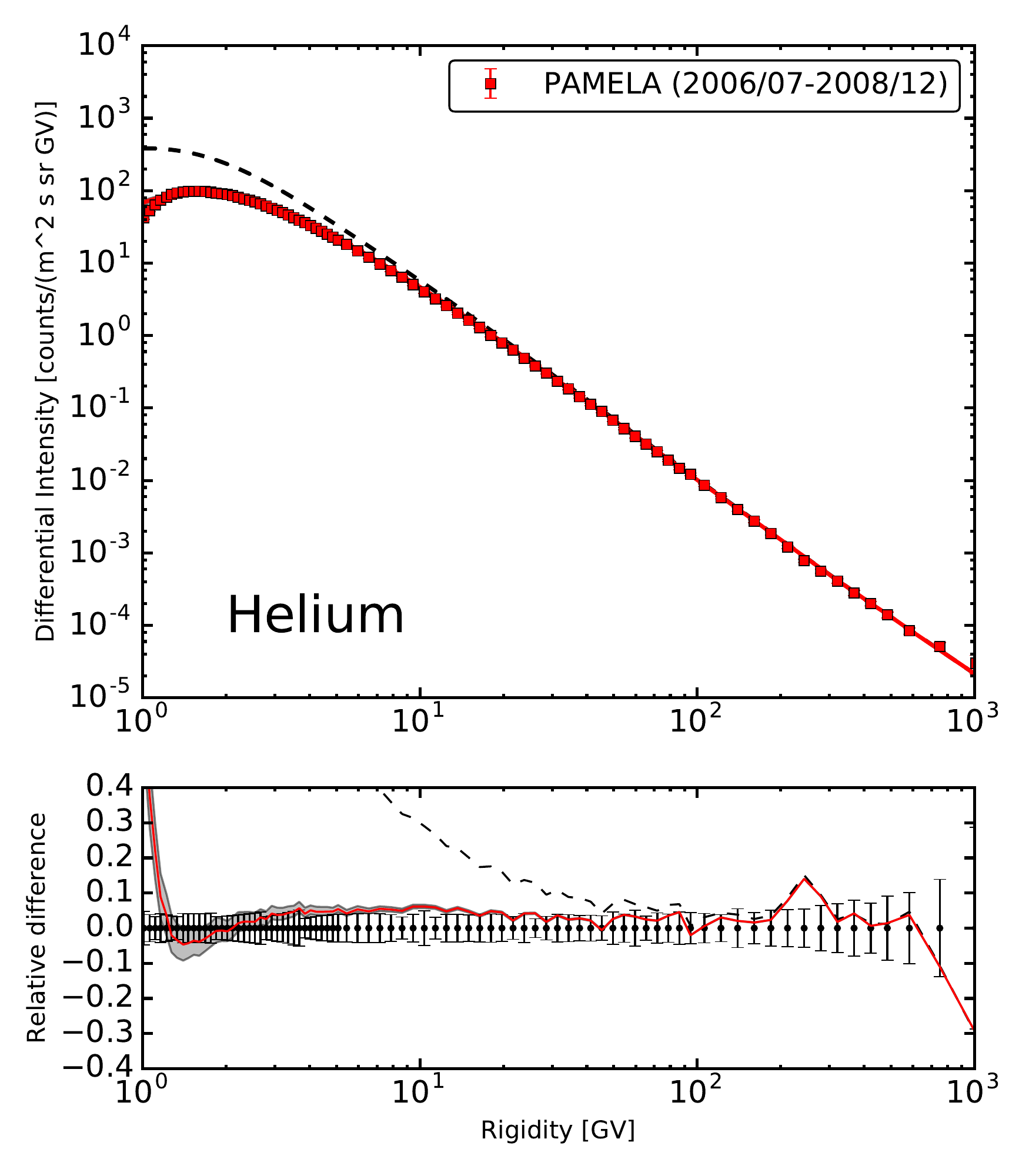}
 }
 \caption{Helium differential intensity for December 2007 (BESS, left) and averaged for 2006-2009 (PAMELA, right). Points represent experimental data, dashed line is the \galprop{} LIS, and solid line is the computed modulated spectrum. The bottom panel is the relative difference between the numerical solution and experimental data.
}
 \label{fig:Helium_Low}
\end{figure*}


\begin{figure*}[!t]
\centerline{
 \includegraphics[width=0.5\textwidth]{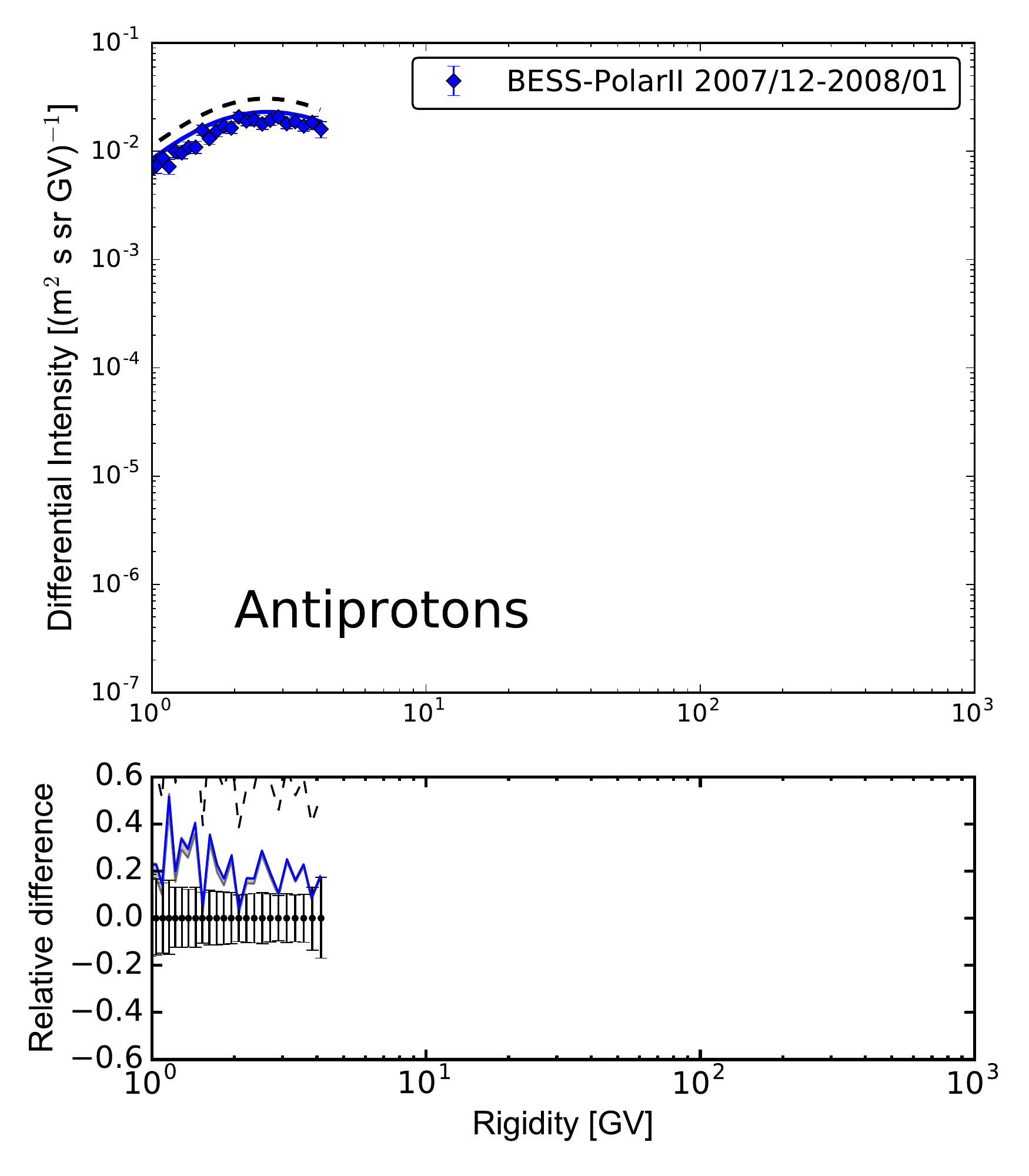}\hfill
 \includegraphics[width=0.5\textwidth]{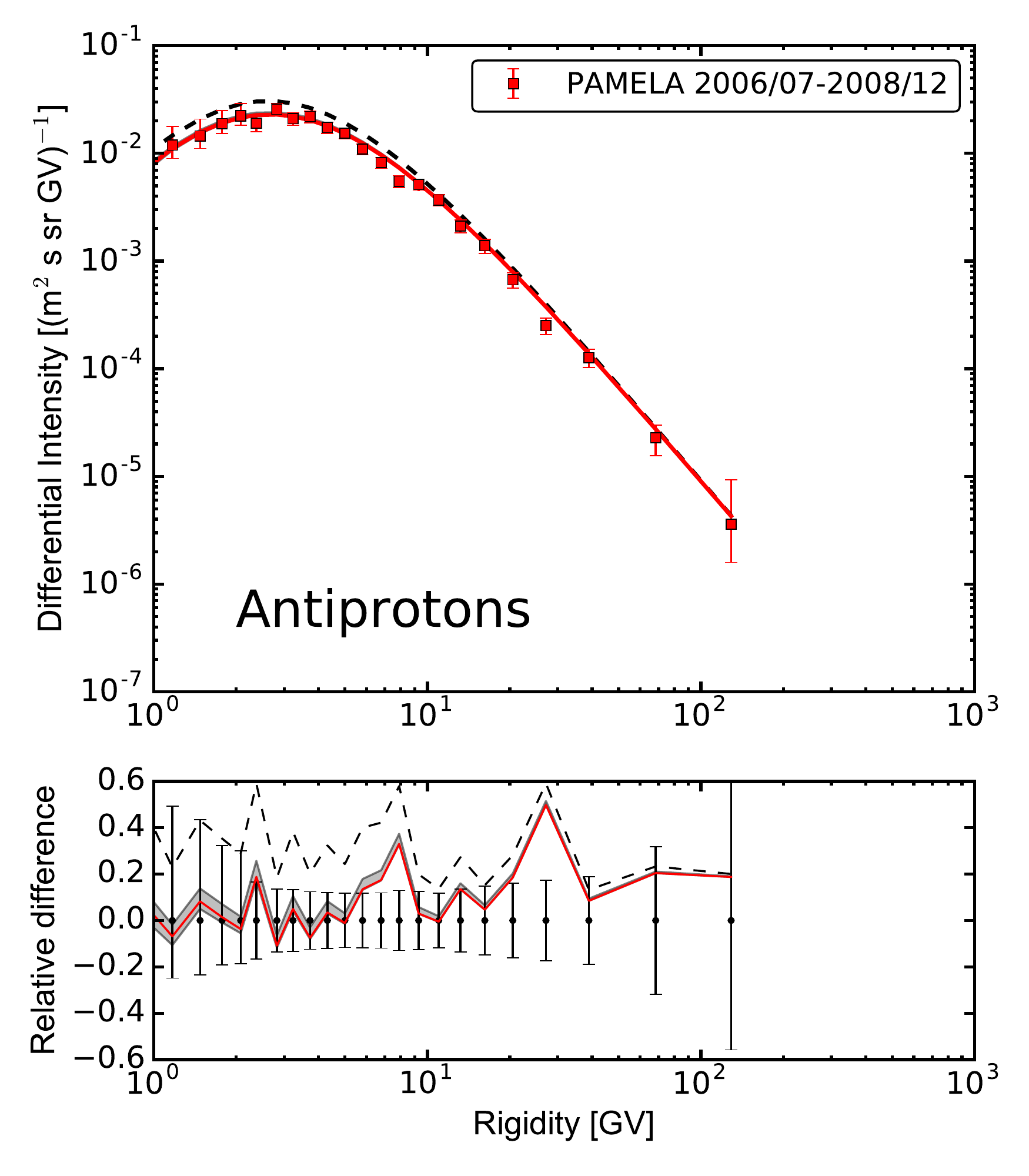}
 }
 \caption{Antiproton differential intensities for December 2007 (BESS, left) and averaged for 2006-2009 (PAMELA, right). Points represent experimental data, dashed line is the \galprop{} LIS, and solid line is the computed modulated spectrum. The bottom panel is the relative difference between the numerical solution and experimental data.}
 \label{fig:Antiproton_Low}
\end{figure*}

\subsubsection{Low Solar Activity} \label{Sect::LowSolar}

During the solar minimum period, the HMF forms a regular structure thus requiring an inclusion of the magnetic drift effects \citep[see e.g.][]{Jokipii77,JokipiiKopriva1979,Potgieter85,BoellaEtAl2001,Strauss2011,DellaTorre2012,DellaTorre2013AdvAstro,ICRC13_DellaTorre}. A low solar activity period between cycles 23 and 24 was recently studied by the PAMELA instrument \citep[see e.g.][]{PamelaProt2013}. Previous missions, AMS-01 mission~\citep[June 1998,][]{AMS01_prot} on the space shuttle, and BESS instrument \citep{bess_prot,BESS2007_Abe_2016}, sampled a few short time periods during solar cycle 23. As discussed in the previous Section, the solar minimum between cycles 23 and 24 was characterized by a negative HMF polarity ($A<0$) that results in a more uniform latitudinal distribution in the inner part of the heliosphere for positively charged CR species. The AMS-01 mission was launched in June 1998 during the solar minimum with
the positive HMF polarity ($A>0$) \citep{AMS01_prot}.

For both solar minima we found that $g_{\rm low}=0.3$ in Eq.~\eqref{EQ::KparActual} leads to an overall agreement between the simulations and data. In Figure \ref{fig:Proton_Low} we show a comparison between experimental data and modulated spectrum calculated with HelMod for June 1998 (AMS-01) and December 2007 (PAMELA). HelMod description of particle propagation takes into account particle type and electrical charge. Note that in Eq.~(\ref{EQ::FPE}) the drift term is the only one that is affected by the charge sign, while all other propagation terms are charge-symmetrical and can be expressed as a function of particle rigidity.

In Figure~\ref{fig:Helium_Low}, the differential helium intensity from BESS-Polar II and PAMELA experiments are compared to the LIS modulated with HelMod. The modulated differential intensity is found in good agreement with the experimental data at all rigidities above 1.5--2 GV, although a deviation relative to PAMELA data is observed at low rigidities.

\begin{figure*}[!tb]
\centerline{
\includegraphics[width=0.72\textwidth]{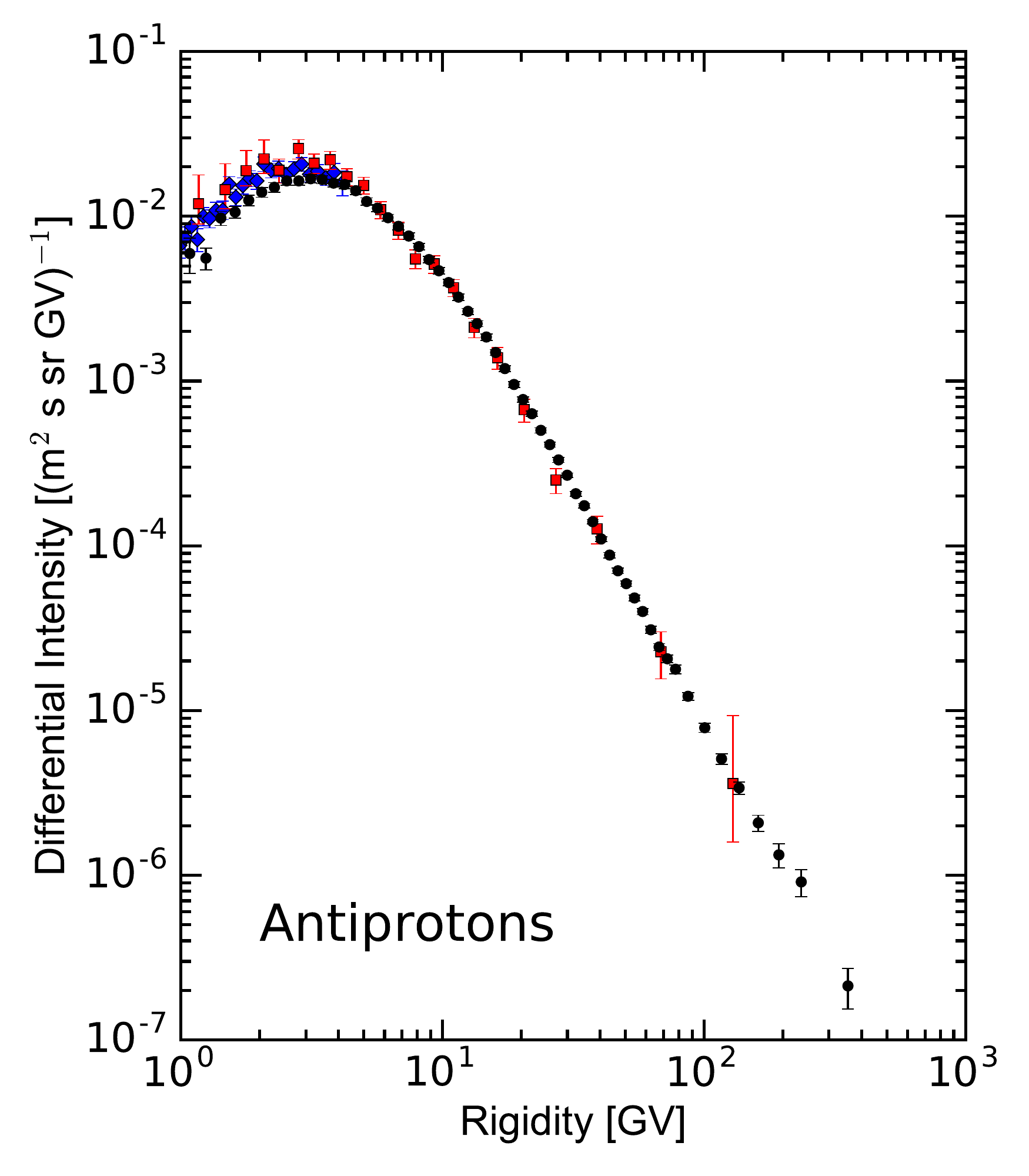}
  }
\caption{Antiproton differential intensity as measured by AMS-02 (black points), BESS-Polar II (red squares), and PAMELA (blue diamonds).}
\label{fig:antip_ExperAll}
\end{figure*}

BESS antiproton data are systematically lower than the antiproton flux calculated with \galprop{}-HelMod (Figure \ref{fig:Antiproton_Low}, left), while no such a discrepancy is observed when calculations are compared to the PAMELA data (Figure \ref{fig:Antiproton_Low}, right). A comparison in Figure \ref{fig:antip_ExperAll} shows that the BESS data points are lower than PAMELA data by an average $\sim$20\% despite the fact that the heliospheric conditions were similar. This may partially explain a slight systematic disagreement of the calculated spectra with BESS-Polar II data.

\begin{figure*}[tbph]
\centerline{
 \includegraphics[width=0.5\textwidth]{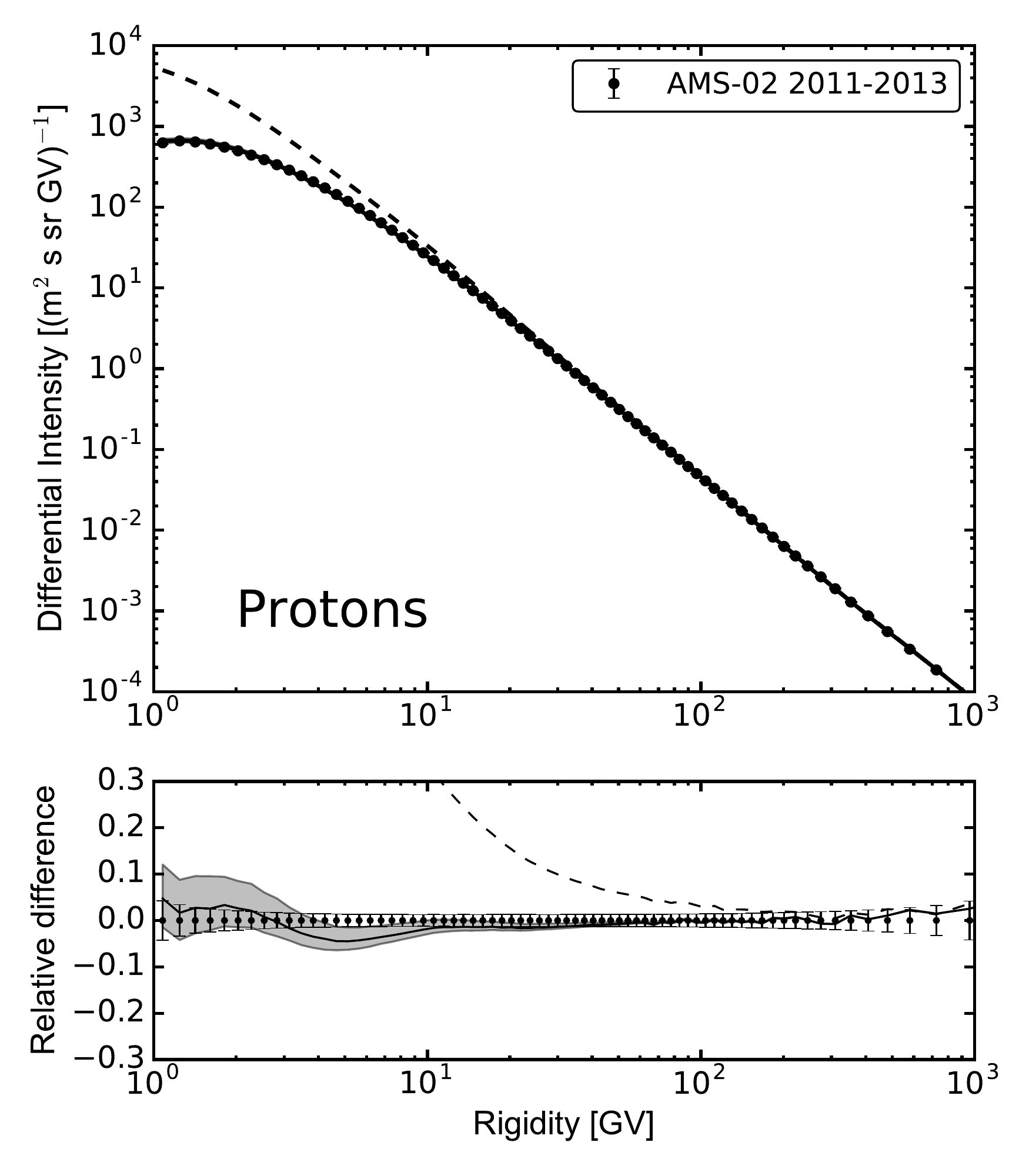}\hfill
 \includegraphics[width=0.5\textwidth]{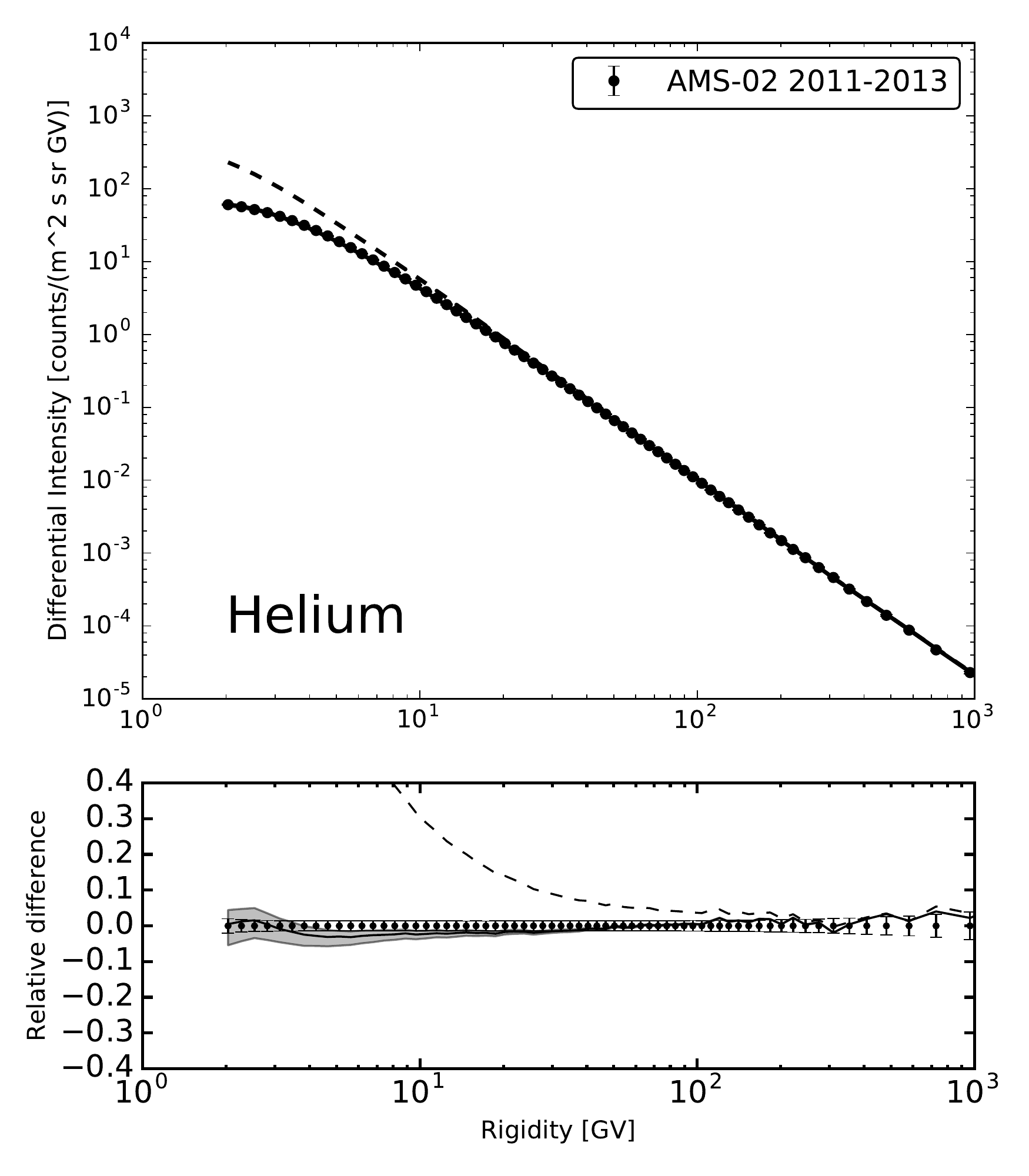}
 }
 \caption{The differential intensities of CR protons (left) and helium (right) measured by AMS-02. Points represent experimental data, dashed line is the \galprop{} LIS, and solid line is the computed modulated spectrum. The bottom panel is the relative difference between the numerical solution and experimental data.}
 \label{fig:pHe_High}
\end{figure*}

\pagebreak[3]
\subsubsection{High Solar Activity}\label{Sect::HighSolar}

High solar activity periods are challenging from the viewpoint of theory of the heliospheric transport. The high frequency of solar events disturbs the interplanetary medium and disrupts the HMF that became difficult to model. Therefore, continuous high precision measurements of CR flux by AMS-02 during a high solar activity period are invaluable. Besides, AMS-02 provides an unique data-set integrated over 3 years \citep{2015PhRvL.114q1103A,2015PhRvL.115u1101A} of observations during the solar activity peak of cycle 24. Previously, BESS instrument \citep{bess_prot} has measured CR protons for a month during the peak of cycle 23.

Not surprisingly, periods of active Sun require some additional refinements of the HelMod parameters. An important change is the lack of regular structure of the HMF that completely suppresses the charge-sign dependence related to the magnetic drift process. As described in Section \ref{Sect::Helmod}, the used description for the magnetic drift velocity, originally developed by \citet{Potgieter85}, already accounts for a reduction of drift transport as solar activity increases, achieved by relating the drift velocity with the tilt angle of the neutral sheet, $\alpha_t$. Similarly to other works \citep[see e.g.,][]{Potgieter_2008}, we introduced an additional correction factor that suppresses any drift velocity during the solar maximum. The second crucial point in describing the high solar activity is related to the diffusion tensor. We found that the best agreement between data and simulations is achieved when we use $g_{\rm low}=0$ in Eq.~\eqref{EQ::KparActual}. This implies that the description of the
magnetic
field turbulence is significantly modified relative to the periods of low solar activity. The described model allows the average proton and helium spectra measured by BESS and AMS-02 during high solar activity periods (see Figure \ref{fig:pHe_High}) to be reproduced reasonably well.

\begin{figure}[tb!]
\centerline{
\includegraphics[width=0.49\textwidth]{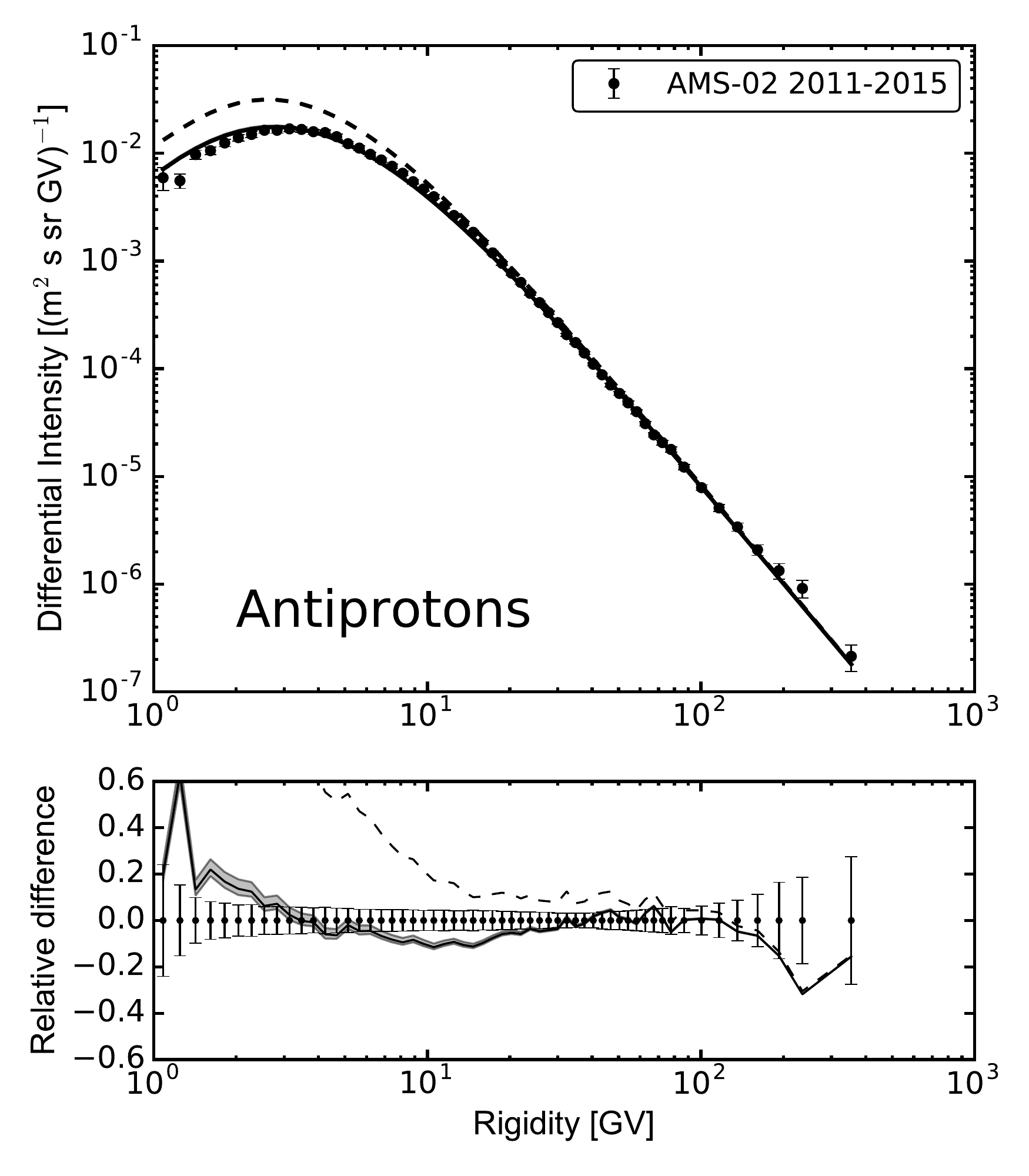}
}
\caption{The differential intensity of CR antiprotons measured by AMS-02. Points represent experimental data, dashed line is the \galprop{} LIS, and solid line is the computed modulated spectrum. The bottom panel is the relative difference between the numerical solution and experimental data. The antiproton calculations include a contribution from nuclei through nickel.}
\label{fig:antip_High}
\end{figure}

\begin{figure}[tbh!]
\centerline{
\includegraphics[width=0.49\textwidth]{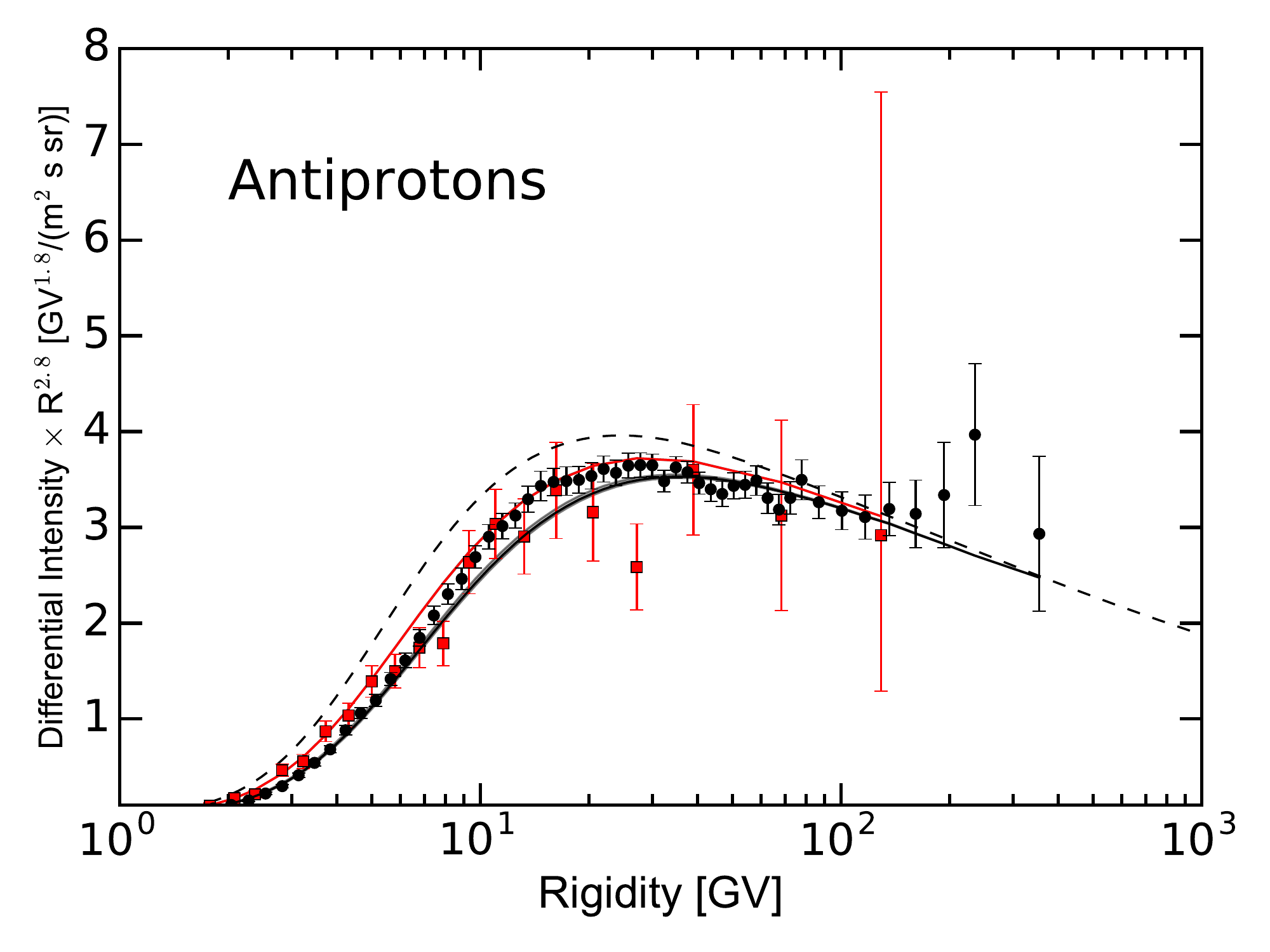}
}
\caption{The differential intensities of CR antiprotons measured by PAMELA (red squares) and AMS-02 (black round). The antiproton LIS is shown by the dashed line. Solid lines correspond to the modulated spectra with HelMod: the lower one in black corresponds to that for AMS-02, the upper in red corresponds to that for PAMELA.
\label{fig:antip_High_r27}}
\end{figure}

In Figure \ref{fig:antip_High} antiproton calculations are compared to the AMS-02 measurements corresponding to a period of high solar activity. The antiproton calculations include a contribution from CR nuclei through nickel. The apparent discrepancy at the lowest rigidity could be due to the irregular behavior of the data points that may indicate some systematic effects. At higher rigidities, between $\sim$7--20 GV, the \galprop{}-HelMod spectrum appears slightly lower than the data (see also Figure \ref{fig:antip_High_r27}), but the discrepancy does not exceed one standard deviation once the HelMod and AMS-02 errors are taken into account. Moreover, inclusion of a contribution of heavier nuclei through nickel improves the agreement with data. At rigidities $\sim$30--100 GV the calculated antiproton flux is slightly higher than the AMS-02 data, but the scattering of the data points in this energy range, clearly visible in Figure~\ref{fig:antip_High_r27}, indicates some additional systematics.


\subsubsection{Ascending and Descending phase}\label{intermediate}

Good quality data available from AMS-02 and PAMELA allowed the CR flux to be continuously observed during the periods of low and high solar activity. However, CR observations during the intermediate activity periods are only available for cycle 23,  these are BESS-1999 \citep{bess_prot} and BESS-Polar I \citep{BESS2007_Abe_2016}. PAMELA data is available until the beginning of 2010, while AMS-02 provides only average values for a period from 2011 to 2013. Therefore, the transitional period is the least studied.

To discriminate between two regimes, we divide the parameter dataset using the tilt angle of the neutral current sheet averaged over the previous year $\bar{\alpha_t}$: low activity occurs if $\bar{\alpha_t} \le50\degree$, and high activity elsewhere. Besides, as described in Section \ref{Sect::Helmod}, the transition periods between low and high activity are treated using a smooth function in the diffusion tensor Eq.~(\ref{EQ::KparActual}) that connects $g_{\rm low}=0$ for high activity with $g_{\rm low}=0.3$ for low activity periods. This parameterization allows the data from BESS-1999 and BESS-2004 to be reproduced fairly well together with the Ulysses normalized counting rate during the transition periods (Figure \ref{fig:OutEcliptic}). The analysis presented in this work also accounts for possible errors in the parameter values determined for the intermediate activity periods.

\section{Conclusions}

One hundred years after the discovery of CRs, the unprecedented precision of AMS-02 instrument and its broad energy coverage promise solutions for many long-standing astrophysical puzzles. However, we are just in the beginning of a fascinating journey. Once the spectra of all elements through iron measured with a few per cent accuracy up to several TeV/nucleon are released, they can be used to identify the origins of CRs and their propagation history, provide new insights into the properties of the interstellar medium, and may reveal new phenomena.
These data would pose a challenge to the theoretical models, but, on the other hand, would also drive us to an ultimate solution. The \galprop{}-HelMod framework described in this paper is providing an example of a self-consistent and concise description of CR propagation from the Galactic scale down to the inner heliosphere. Elimination of the uncertainties in the astrophysical backgrounds would, in turn, enable us to search for traces of exotic physics.

The proton and helium LIS derived in the current work, allow all the data for solar cycles 23 and 24 to be successfully reproduced within a single framework. This includes a fully realistic and exhaustive description of the relevant CR physics. The proposed LIS accommodate both the very low energy interstellar CR spectra measured by Voyager 1 and the higher energy observations at Earth publicly released by BESS, PAMELA, AMS-01, and AMS-02. Given their high precision, recent AMS-02 antiproton data pose a serious challenge for propagation models. However, the proposed models provide a good description in the whole energy range with the maximal deviations of the order of one standard error thus illustrating a significant potential of the combined \galprop{}-HelMod framework.


\acknowledgements
We wish to specially thank Pavol Bobik, Giuliano Boella, Karel Kudela, Marian Putis and Mario Zannoni for their support to the HelMod code and many useful suggestions.
This work is supported by ASI (Agenzia Spaziale Italiana) under contract ASI-INFN I/002/13/0 and by ESA (European Space Agency) contract 4000116146/16/NL/HK. Igor Moskalenko, Elena Orlando, Troy Porter acknowledge support from NASA Grants Nos.~NNX13AC47G and NNX17AB48G. Elena Orlando additionally acknowledges support from NASA Grants Nos.~NNX16AF27G and NNX15AU79G.
Sergey Ostapchenko acknowledges support from Deutsche Forschungsgemeinschaft project OS 481/1.

\bibliography{bibliography,proposal_moskalenko,imos,imos_icrc2015,biblio_masi,HM_RV}


\clearpage

\appendix
\section{Supplementary Material}


\begin{figure*}[!h]
\centerline{
\includegraphics[width=0.5\textwidth]{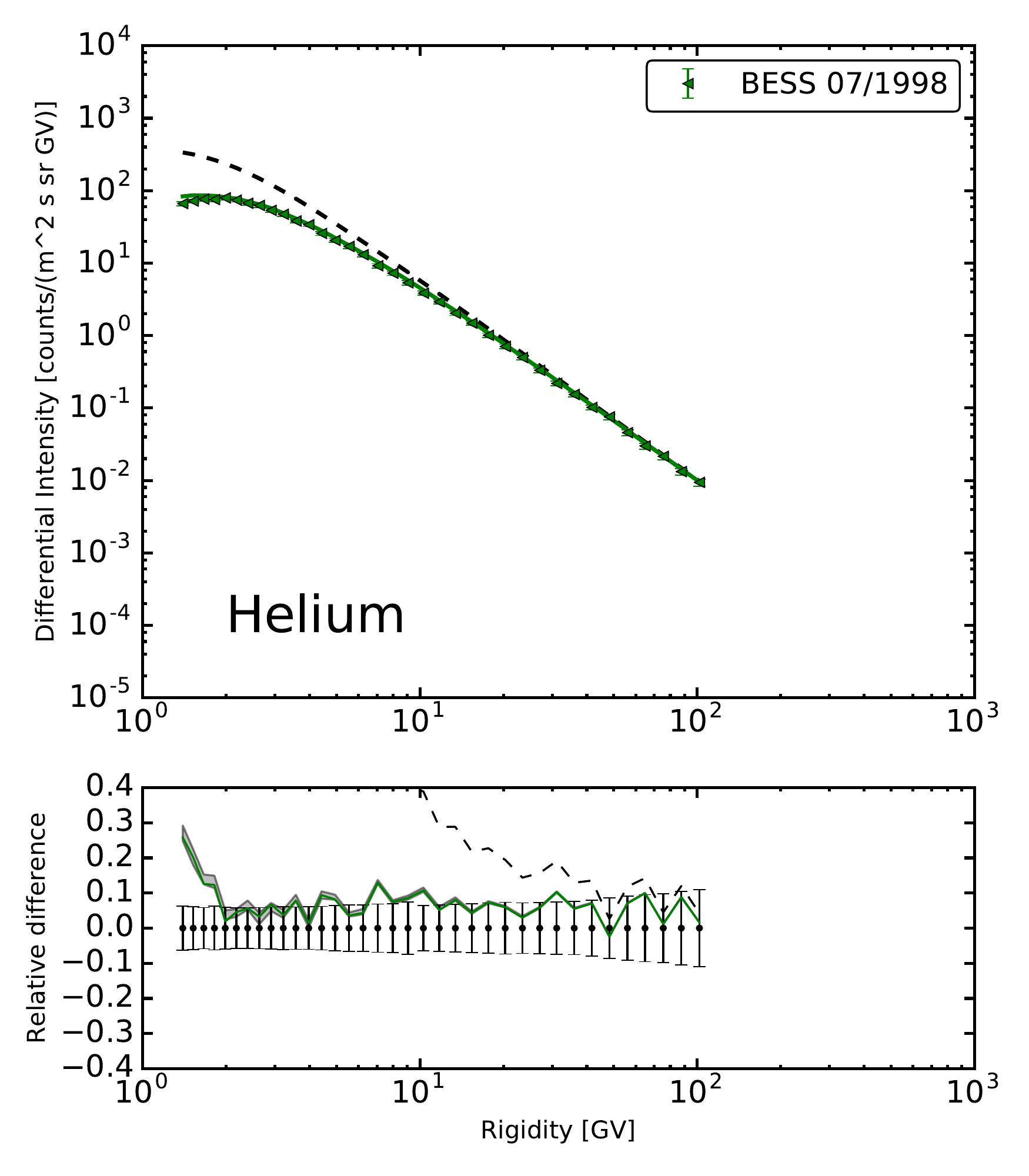}\hfill
\includegraphics[width=0.5\textwidth]{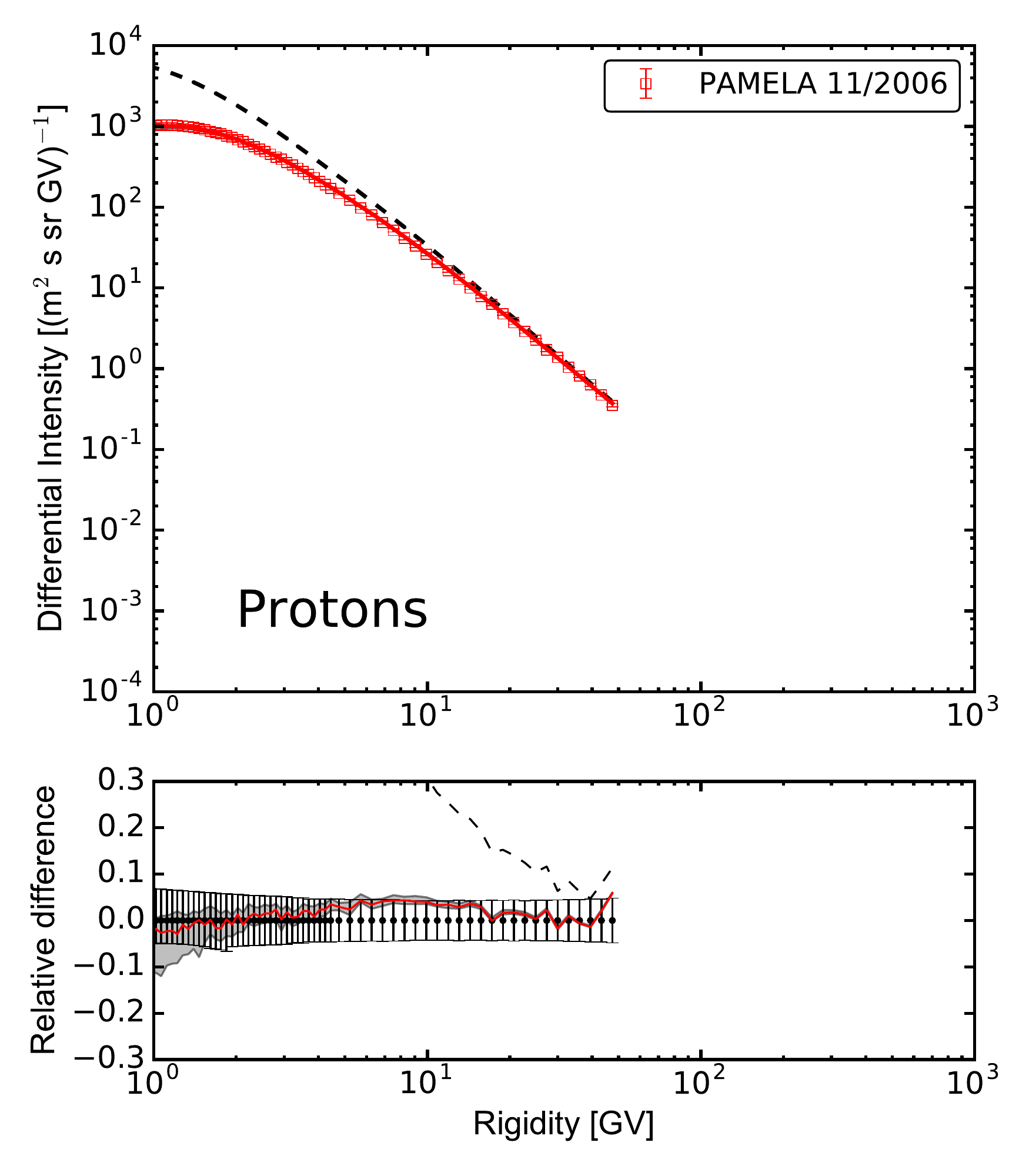}
}
\caption{Left: Helium differential intensities for July 1998 (BESS). Right: Proton differential intensities for December 2006 (PAMELA).
Points represent experimental measurements, dashed
line is the GALPROP LIS and solid line is the computed modulated spectrum. The bottom
panel is the relative difference between numerical solution and experimental data.}
\label{fig:HelMod_Helium_BESS1998_Rigi_c}
\end{figure*}

\begin{figure*}[!h]
\centerline{
\includegraphics[width=0.5\textwidth]{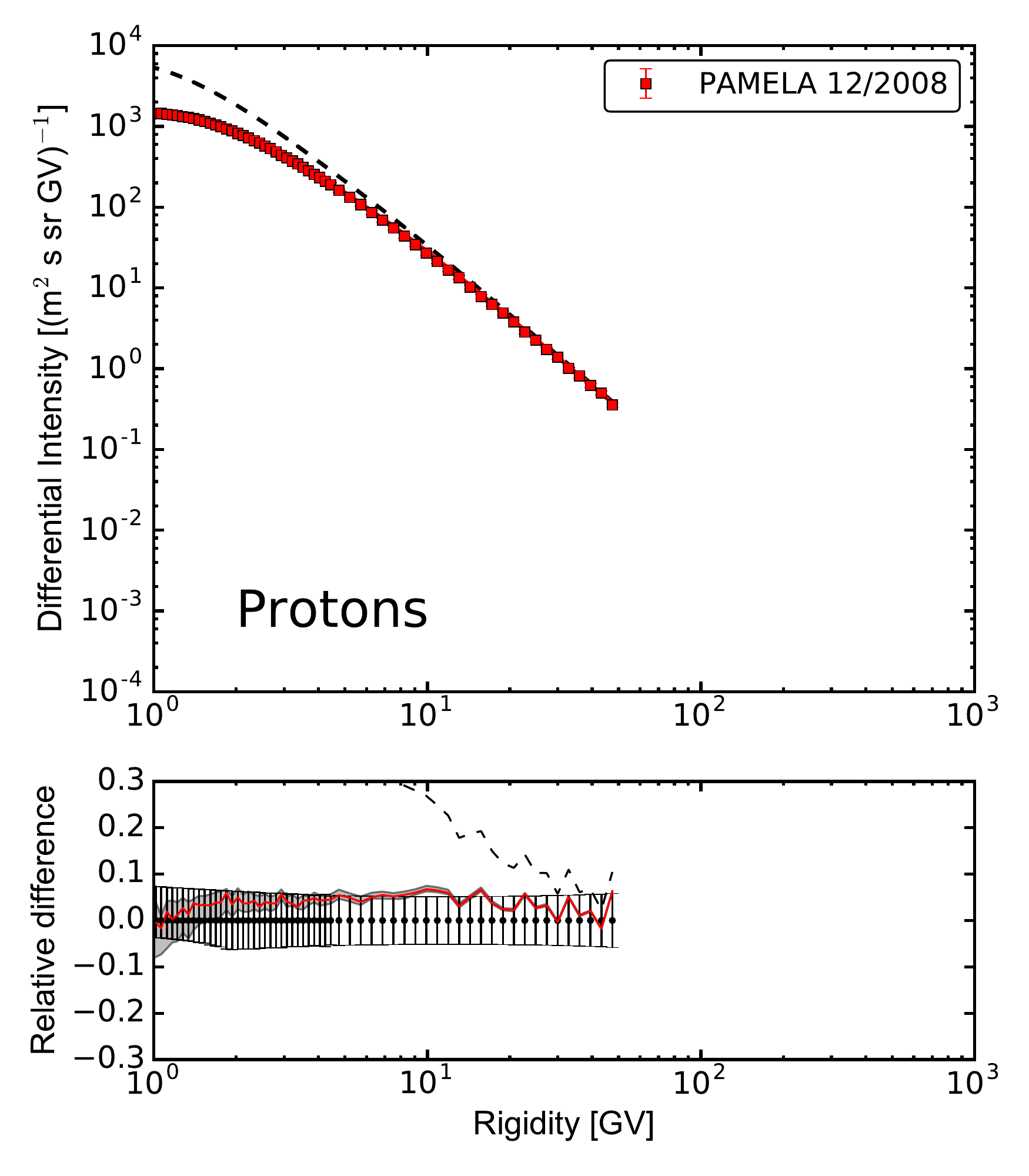}\hfill
\includegraphics[width=0.5\textwidth]{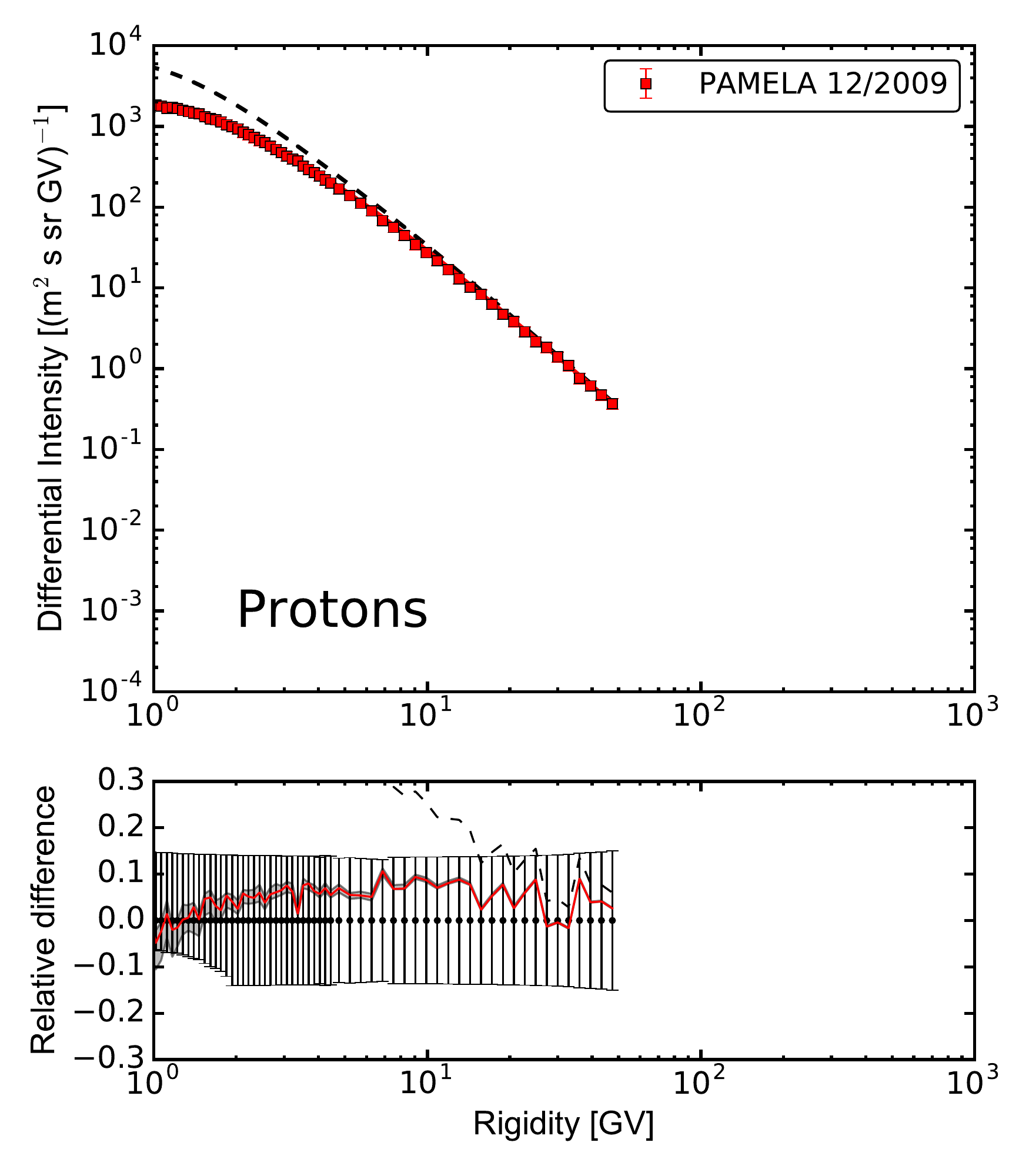}
}
\caption{Left: Proton differential intensities for December 2008 (PAMELA). Right: Proton differential intensities for December 2009 (PAMELA).
See Figure \ref{fig:HelMod_Helium_BESS1998_Rigi_c} for legend.}
\label{fig:HelMod_Proton_PAMELA08_Rigi_c}
\end{figure*}

\begin{deluxetable}{cccccccccc}
\tablecolumns{10}
\tablewidth{0mm}
\tablecaption{Proton LIS\label{Tbl-ProtonLIS}}
\tablehead{
\colhead{Rigidity} & \colhead{Differential} &
\colhead{Rigidity} & \colhead{Differential} &
\colhead{Rigidity} & \colhead{Differential} &
\colhead{Rigidity} & \colhead{Differential} &
\colhead{Rigidity} & \colhead{Differential}
\\
\colhead{GV} & \colhead{Intensity\tablenotemark{a}} &
\colhead{GV} & \colhead{Intensity\tablenotemark{a}} &
\colhead{GV} & \colhead{Intensity\tablenotemark{a}} &
\colhead{GV} & \colhead{Intensity\tablenotemark{a}} &
\colhead{GV} & \colhead{Intensity\tablenotemark{a}} 
}
\startdata
4.333e-02&2.971e+02  &2.814e-01&4.265e+03  &2.473e+00&1.180e+03  &7.145e+01&1.226e-01  & 2.914e+03&4.554e-06\\
4.482e-02&3.772e+02  &2.913e-01&4.411e+03  &2.600e+00&1.054e+03  &7.639e+01&1.014e-01  & 3.118e+03&3.799e-06\\
4.636e-02&4.101e+02  &3.016e-01&4.557e+03  &2.736e+00&9.393e+02  &8.167e+01&8.382e-02  & 3.336e+03&3.169e-06\\
4.796e-02&4.325e+02  &3.123e-01&4.703e+03  &2.880e+00&8.337e+02  &8.732e+01&6.927e-02  & 3.570e+03&2.643e-06\\
4.961e-02&4.531e+02  &3.233e-01&4.849e+03  &3.033e+00&7.377e+02  &9.337e+01&5.725e-02  & 3.820e+03&2.204e-06\\
5.132e-02&4.738e+02  &3.348e-01&4.992e+03  &3.197e+00&6.509e+02  &9.984e+01&4.731e-02  & 4.087e+03&1.839e-06\\
5.308e-02&4.953e+02  &3.466e-01&5.134e+03  &3.371e+00&5.726e+02  &1.067e+02&3.909e-02  & 4.373e+03&1.534e-06\\
5.491e-02&5.178e+02  &3.590e-01&5.273e+03  &3.557e+00&5.024e+02  &1.141e+02&3.230e-02  & 4.679e+03&1.279e-06\\
5.680e-02&5.414e+02  &3.718e-01&5.408e+03  &3.755e+00&4.396e+02  &1.221e+02&2.669e-02  & 5.007e+03&1.067e-06\\
5.876e-02&5.660e+02  &3.851e-01&5.539e+03  &3.966e+00&3.838e+02  &1.305e+02&2.205e-02  & 5.357e+03&8.901e-07\\
6.078e-02&5.919e+02  &3.988e-01&5.665e+03  &4.192e+00&3.342e+02  &1.396e+02&1.821e-02  & 5.732e+03&7.424e-07\\
6.288e-02&6.190e+02  &4.132e-01&5.785e+03  &4.432e+00&2.904e+02  &1.493e+02&1.505e-02  & 6.133e+03&6.192e-07\\
6.504e-02&6.474e+02  &4.280e-01&5.899e+03  &4.689e+00&2.518e+02  &1.597e+02&1.243e-02  & 6.562e+03&5.164e-07\\
6.729e-02&6.772e+02  &4.435e-01&6.006e+03  &4.963e+00&2.178e+02  &1.708e+02&1.027e-02  & 7.022e+03&4.307e-07\\
6.960e-02&7.085e+02  &4.596e-01&6.104e+03  &5.256e+00&1.880e+02  &1.827e+02&8.483e-03  & 7.513e+03&3.592e-07\\
7.200e-02&7.412e+02  &4.763e-01&6.194e+03  &5.569e+00&1.620e+02  &1.955e+02&7.007e-03  & 8.039e+03&2.996e-07\\
7.448e-02&7.756e+02  &4.937e-01&6.274e+03  &5.903e+00&1.392e+02  &2.091e+02&5.787e-03  & 8.602e+03&2.499e-07\\
7.705e-02&8.117e+02  &5.117e-01&6.343e+03  &6.260e+00&1.194e+02  &2.237e+02&4.780e-03  & 9.204e+03&2.084e-07\\
7.971e-02&8.495e+02  &5.306e-01&6.402e+03  &6.641e+00&1.022e+02  &2.392e+02&3.948e-03  & 9.848e+03&1.738e-07\\
8.246e-02&8.893e+02  &5.501e-01&6.449e+03  &7.049e+00&8.735e+01  &2.559e+02&3.261e-03  & 1.053e+04&1.449e-07\\
8.530e-02&9.309e+02  &5.705e-01&6.483e+03  &7.484e+00&7.444e+01  &2.738e+02&2.694e-03  & 1.127e+04&1.209e-07\\
8.824e-02&9.747e+02  &5.918e-01&6.504e+03  &7.950e+00&6.331e+01  &2.929e+02&2.226e-03  & 1.206e+04&1.008e-07\\
9.128e-02&1.020e+03  &6.139e-01&6.512e+03  &8.448e+00&5.372e+01  &3.133e+02&1.840e-03  & 1.290e+04&8.410e-08\\
9.443e-02&1.068e+03  &6.370e-01&6.505e+03  &8.981e+00&4.548e+01  &3.352e+02&1.523e-03  & 1.381e+04&7.014e-08\\
9.769e-02&1.119e+03  &6.611e-01&6.484e+03  &9.550e+00&3.842e+01  &3.586e+02&1.262e-03  & 1.477e+04&5.850e-08\\
1.010e-01&1.172e+03  &6.863e-01&6.447e+03  &1.015e+01&3.239e+01  &3.836e+02&1.049e-03  & 1.581e+04&4.878e-08\\
1.045e-01&1.227e+03  &7.126e-01&6.395e+03  &1.081e+01&2.725e+01  &4.104e+02&8.731e-04  & 1.692e+04&4.068e-08\\
1.081e-01&1.286e+03  &7.400e-01&6.328e+03  &1.150e+01&2.288e+01  &4.391e+02&7.275e-04  & 1.810e+04&3.393e-08\\
1.118e-01&1.347e+03  &7.687e-01&6.244e+03  &1.225e+01&1.918e+01  &4.698e+02&6.067e-04  & 1.937e+04&2.829e-08\\
1.157e-01&1.411e+03  &7.987e-01&6.145e+03  &1.304e+01&1.605e+01  &5.026e+02&5.061e-04  & 2.072e+04&2.360e-08\\
1.197e-01&1.477e+03  &8.301e-01&6.030e+03  &1.390e+01&1.341e+01  &5.377e+02&4.223e-04  & 2.218e+04&1.968e-08\\
1.238e-01&1.548e+03  &8.629e-01&5.900e+03  &1.481e+01&1.119e+01  &5.753e+02&3.523e-04  & 2.373e+04&1.641e-08\\
1.281e-01&1.621e+03  &8.974e-01&5.755e+03  &1.578e+01&9.325e+00  &6.155e+02&2.940e-04  & 2.539e+04&1.368e-08\\
1.326e-01&1.698e+03  &9.335e-01&5.595e+03  &1.682e+01&7.762e+00  &6.585e+02&2.453e-04  & 2.717e+04&1.141e-08\\
1.372e-01&1.778e+03  &9.713e-01&5.422e+03  &1.794e+01&6.455e+00  &7.046e+02&2.047e-04  & 2.907e+04&9.519e-09\\
1.419e-01&1.861e+03  &1.011e+00&5.237e+03  &1.913e+01&5.364e+00  &7.538e+02&1.708e-04  & 3.110e+04&7.938e-09\\
1.468e-01&1.948e+03  &1.052e+00&5.039e+03  &2.041e+01&4.454e+00  &8.065e+02&1.425e-04  & 3.328e+04&6.619e-09\\
1.519e-01&2.039e+03  &1.096e+00&4.830e+03  &2.178e+01&3.696e+00  &8.629e+02&1.189e-04  & 3.561e+04&5.520e-09\\
1.572e-01&2.134e+03  &1.142e+00&4.612e+03  &2.324e+01&3.065e+00  &9.233e+02&9.921e-05  & 3.810e+04&4.603e-09\\
1.626e-01&2.232e+03  &1.191e+00&4.385e+03  &2.480e+01&2.541e+00  &9.878e+02&8.277e-05  & 4.077e+04&3.838e-09\\
1.682e-01&2.334e+03  &1.242e+00&4.153e+03  &2.648e+01&2.105e+00  &1.056e+03&6.905e-05  & 4.363e+04&3.201e-09\\
1.741e-01&2.440e+03  &1.296e+00&3.917e+03  &2.827e+01&1.744e+00  &1.130e+03&5.761e-05  & 4.668e+04&2.669e-09\\
1.801e-01&2.550e+03  &1.353e+00&3.681e+03  &3.018e+01&1.444e+00  &1.209e+03&4.806e-05  & 4.995e+04&2.226e-09\\
1.864e-01&2.663e+03  &1.413e+00&3.447e+03  &3.223e+01&1.196e+00  &1.294e+03&4.010e-05  & 5.344e+04&1.856e-09\\
1.929e-01&2.780e+03  &1.476e+00&3.216e+03  &3.442e+01&9.902e-01  &1.385e+03&3.345e-05  & 5.719e+04&1.547e-09\\
1.996e-01&2.901e+03  &1.543e+00&2.989e+03  &3.677e+01&8.195e-01  &1.482e+03&2.791e-05  & 6.119e+04&1.290e-09\\
2.065e-01&3.025e+03  &1.613e+00&2.769e+03  &3.928e+01&6.781e-01  &1.585e+03&2.328e-05  & 6.547e+04&1.076e-09\\
2.137e-01&3.152e+03  &1.688e+00&2.556e+03  &4.197e+01&5.610e-01  &1.696e+03&1.942e-05  & 7.006e+04&8.974e-10\\
2.212e-01&3.283e+03  &1.767e+00&2.351e+03  &4.484e+01&4.641e-01  &1.815e+03&1.620e-05  & 7.496e+04&7.482e-10\\
2.289e-01&3.416e+03  &1.851e+00&2.154e+03  &4.792e+01&3.839e-01  &1.942e+03&1.351e-05  & 8.021e+04&6.239e-10\\
2.369e-01&3.553e+03  &1.940e+00&1.966e+03  &5.121e+01&3.175e-01  &2.078e+03&1.127e-05  & 8.582e+04&5.202e-10\\
2.452e-01&3.691e+03  &2.034e+00&1.789e+03  &5.473e+01&2.625e-01  &2.223e+03&9.406e-06  & 9.183e+04&4.337e-10\\
2.538e-01&3.832e+03  &2.134e+00&1.621e+03  &5.849e+01&2.171e-01  &2.379e+03&7.846e-06  & 9.826e+04&3.609e-10\\
2.627e-01&3.975e+03  &2.240e+00&1.463e+03  &6.252e+01&1.795e-01  &2.545e+03&6.545e-06  & 1.051e+05&2.844e-10\\
2.719e-01&4.120e+03  &2.353e+00&1.316e+03  &6.683e+01&1.484e-01  &2.723e+03&5.460e-06  & \nodata & \nodata
\enddata
\tablenotetext{a}{Differential Intensity units: (m$^2$ s sr GV)$^{-1}$.}
\end{deluxetable}

\begin{deluxetable}{cccccccccc}
\tablecolumns{10}
\tablewidth{0mm}
\tablecaption{Helium LIS\label{Tbl-HeliumLIS}}
\tablehead{
\colhead{Rigidity} & \colhead{Differential} &
\colhead{Rigidity} & \colhead{Differential} &
\colhead{Rigidity} & \colhead{Differential} &
\colhead{Rigidity} & \colhead{Differential} &
\colhead{Rigidity} & \colhead{Differential}
\\
\colhead{GV} & \colhead{Intensity\tablenotemark{a}} &
\colhead{GV} & \colhead{Intensity\tablenotemark{a}} &
\colhead{GV} & \colhead{Intensity\tablenotemark{a}} &
\colhead{GV} & \colhead{Intensity\tablenotemark{a}} &
\colhead{GV} & \colhead{Intensity\tablenotemark{a}} 
}
\startdata
8.634e-02&2.144e+01  &5.609e-01&2.506e+02  &4.937e+00&3.551e+01  &1.428e+02&3.842e-03  &5.829e+03&2.188e-07 \\
8.932e-02&2.621e+01  &5.807e-01&2.583e+02  &5.191e+00&3.131e+01  &1.527e+02&3.194e-03  &6.236e+03&1.835e-07 \\
9.239e-02&2.808e+01  &6.011e-01&2.660e+02  &5.462e+00&2.755e+01  &1.633e+02&2.655e-03  &6.673e+03&1.539e-07 \\
9.557e-02&2.945e+01  &6.224e-01&2.738e+02  &5.750e+00&2.418e+01  &1.746e+02&2.206e-03  &7.140e+03&1.290e-07 \\
9.886e-02&3.076e+01  &6.444e-01&2.816e+02  &6.057e+00&2.119e+01  &1.867e+02&1.834e-03  &7.640e+03&1.082e-07 \\
1.022e-01&3.211e+01  &6.672e-01&2.895e+02  &6.384e+00&1.852e+01  &1.996e+02&1.523e-03  &8.174e+03&9.078e-08 \\
1.057e-01&3.351e+01  &6.909e-01&2.974e+02  &6.733e+00&1.616e+01  &2.135e+02&1.265e-03  &8.746e+03&7.613e-08 \\
1.094e-01&3.498e+01  &7.155e-01&3.052e+02  &7.104e+00&1.407e+01  &2.283e+02&1.051e-03  &9.359e+03&6.384e-08 \\
1.132e-01&3.651e+01  &7.410e-01&3.129e+02  &7.500e+00&1.222e+01  &2.441e+02&8.730e-04  &1.001e+04&5.354e-08 \\
1.171e-01&3.812e+01  &7.675e-01&3.207e+02  &7.923e+00&1.060e+01  &2.611e+02&7.251e-04  &1.071e+04&4.489e-08 \\
1.211e-01&3.979e+01  &7.950e-01&3.287e+02  &8.373e+00&9.178e+00  &2.793e+02&6.024e-04  &1.146e+04&3.765e-08 \\
1.253e-01&4.155e+01  &8.235e-01&3.369e+02  &8.854e+00&7.927e+00  &2.987e+02&5.008e-04  &1.226e+04&3.157e-08 \\
1.296e-01&4.338e+01  &8.532e-01&3.437e+02  &9.368e+00&6.834e+00  &3.195e+02&4.171e-04  &1.312e+04&2.647e-08 \\
1.340e-01&4.530e+01  &8.840e-01&3.494e+02  &9.916e+00&5.881e+00  &3.417e+02&3.484e-04  &1.404e+04&2.220e-08 \\
1.387e-01&4.731e+01  &9.161e-01&3.541e+02  &1.050e+01&5.050e+00  &3.655e+02&2.917e-04  &1.502e+04&1.861e-08 \\
1.434e-01&4.941e+01  &9.494e-01&3.578e+02  &1.112e+01&4.329e+00  &3.910e+02&2.447e-04  &1.607e+04&1.561e-08 \\
1.484e-01&5.161e+01  &9.840e-01&3.605e+02  &1.179e+01&3.704e+00  &4.182e+02&2.054e-04  &1.720e+04&1.308e-08 \\
1.535e-01&5.392e+01  &1.020e+00&3.623e+02  &1.250e+01&3.163e+00  &4.473e+02&1.724e-04  &1.840e+04&1.097e-08 \\
1.588e-01&5.633e+01  &1.057e+00&3.630e+02  &1.327e+01&2.696e+00  &4.785e+02&1.447e-04  &1.969e+04&9.202e-09 \\
1.643e-01&5.885e+01  &1.096e+00&3.627e+02  &1.408e+01&2.294e+00  &5.119e+02&1.215e-04  &2.107e+04&7.716e-09 \\
1.699e-01&6.149e+01  &1.137e+00&3.614e+02  &1.495e+01&1.949e+00  &5.476e+02&1.020e-04  &2.255e+04&6.470e-09 \\
1.758e-01&6.425e+01  &1.179e+00&3.590e+02  &1.588e+01&1.654e+00  &5.858e+02&8.562e-05  &2.412e+04&5.424e-09 \\
1.819e-01&6.714e+01  &1.223e+00&3.555e+02  &1.688e+01&1.400e+00  &6.267e+02&7.187e-05  &2.581e+04&4.548e-09 \\
1.881e-01&7.017e+01  &1.270e+00&3.510e+02  &1.795e+01&1.184e+00  &6.704e+02&6.032e-05  &2.762e+04&3.813e-09 \\
1.946e-01&7.333e+01  &1.318e+00&3.455e+02  &1.908e+01&1.000e+00  &7.172e+02&5.062e-05  &2.955e+04&3.197e-09 \\
2.014e-01&7.663e+01  &1.368e+00&3.390e+02  &2.030e+01&8.443e-01  &7.673e+02&4.248e-05  &3.162e+04&2.681e-09 \\
2.083e-01&8.009e+01  &1.420e+00&3.315e+02  &2.160e+01&7.115e-01  &8.209e+02&3.565e-05  &3.384e+04&2.247e-09 \\
2.155e-01&8.370e+01  &1.475e+00&3.230e+02  &2.300e+01&5.989e-01  &8.782e+02&2.992e-05  &3.621e+04&1.884e-09 \\
2.229e-01&8.746e+01  &1.532e+00&3.137e+02  &2.449e+01&5.037e-01  &9.396e+02&2.510e-05  &3.874e+04&1.580e-09 \\
2.306e-01&9.140e+01  &1.592e+00&3.035e+02  &2.608e+01&4.233e-01  &1.005e+03&2.106e-05  &4.145e+04&1.324e-09 \\
2.386e-01&9.550e+01  &1.655e+00&2.926e+02  &2.778e+01&3.554e-01  &1.075e+03&1.767e-05  &4.436e+04&1.110e-09 \\
2.469e-01&9.977e+01  &1.720e+00&2.810e+02  &2.961e+01&2.982e-01  &1.150e+03&1.483e-05  &4.746e+04&9.312e-10 \\
2.554e-01&1.042e+02  &1.789e+00&2.688e+02  &3.156e+01&2.500e-01  &1.231e+03&1.244e-05  &5.078e+04&7.807e-10 \\
2.642e-01&1.088e+02  &1.861e+00&2.562e+02  &3.364e+01&2.095e-01  &1.317e+03&1.044e-05  &5.434e+04&6.545e-10 \\
2.734e-01&1.136e+02  &1.936e+00&2.433e+02  &3.587e+01&1.754e-01  &1.409e+03&8.762e-06  &5.814e+04&5.487e-10 \\
2.828e-01&1.186e+02  &2.016e+00&2.302e+02  &3.826e+01&1.468e-01  &1.507e+03&7.352e-06  &6.221e+04&4.600e-10 \\
2.926e-01&1.238e+02  &2.099e+00&2.170e+02  &4.082e+01&1.228e-01  &1.613e+03&6.168e-06  &6.657e+04&3.856e-10 \\
3.027e-01&1.291e+02  &2.187e+00&2.037e+02  &4.355e+01&1.026e-01  &1.725e+03&5.174e-06  &7.123e+04&3.233e-10 \\
3.132e-01&1.347e+02  &2.279e+00&1.905e+02  &4.647e+01&8.578e-02  &1.846e+03&4.341e-06  &7.621e+04&2.710e-10 \\
3.241e-01&1.404e+02  &2.376e+00&1.775e+02  &4.960e+01&7.163e-02  &1.975e+03&3.642e-06  &8.155e+04&2.272e-10 \\
3.353e-01&1.463e+02  &2.478e+00&1.647e+02  &5.295e+01&5.979e-02  &2.113e+03&3.055e-06  &8.726e+04&1.905e-10 \\
3.470e-01&1.524e+02  &2.585e+00&1.523e+02  &5.653e+01&4.989e-02  &2.261e+03&2.563e-06  &9.336e+04&1.597e-10 \\
3.590e-01&1.586e+02  &2.699e+00&1.402e+02  &6.036e+01&4.161e-02  &2.419e+03&2.150e-06  &9.990e+04&1.339e-10 \\
3.715e-01&1.650e+02  &2.818e+00&1.287e+02  &6.446e+01&3.470e-02  &2.589e+03&1.804e-06  &1.068e+05&1.122e-10 \\
3.844e-01&1.715e+02  &2.945e+00&1.176e+02  &6.884e+01&2.892e-02  &2.770e+03&1.513e-06  &1.143e+05&9.410e-11 \\
3.978e-01&1.782e+02  &3.078e+00&1.072e+02  &7.353e+01&2.410e-02  &2.964e+03&1.269e-06  &1.223e+05&7.889e-11 \\
4.116e-01&1.850e+02  &3.219e+00&9.741e+01  &7.855e+01&2.008e-02  &3.171e+03&1.064e-06  &1.309e+05&6.613e-11 \\
4.260e-01&1.919e+02  &3.369e+00&8.817e+01  &8.393e+01&1.672e-02  &3.393e+03&8.932e-07  &1.401e+05&5.544e-11 \\
4.408e-01&1.989e+02  &3.527e+00&7.952e+01  &8.967e+01&1.392e-02  &3.630e+03&7.493e-07  &1.499e+05&4.647e-11 \\
4.562e-01&2.061e+02  &3.694e+00&7.150e+01  &9.582e+01&1.159e-02  &3.884e+03&6.285e-07  &1.604e+05&3.895e-11 \\
4.722e-01&2.133e+02  &3.872e+00&6.408e+01  &1.024e+02&9.648e-03  &4.156e+03&5.271e-07  &1.716e+05&3.263e-11 \\
4.887e-01&2.207e+02  &4.060e+00&5.726e+01  &1.094e+02&8.028e-03  &4.447e+03&4.421e-07  &1.836e+05&2.733e-11 \\
5.058e-01&2.281e+02  &4.260e+00&5.101e+01  &1.169e+02&6.679e-03  &4.758e+03&3.708e-07  &1.965e+05&2.285e-11 \\
5.235e-01&2.355e+02  &4.472e+00&4.533e+01  &1.250e+02&5.556e-03  &5.091e+03&3.110e-07  &2.102e+05&1.831e-11 \\   
5.419e-01&2.431e+02  &4.697e+00&4.017e+01  &1.336e+02&4.620e-03  &5.447e+03&2.608e-07  & \nodata&\nodata
\enddata
\tablenotetext{a}{Differential Intensity units: (m$^2$ s sr GV)$^{-1}$.}
\end{deluxetable}

\begin{deluxetable}{cccccccccc}
\tablecolumns{10}
\tablewidth{0mm}
\tablecaption{Antiproton LIS\label{Tbl-AntiprotonLIS}}
\tablehead{
\colhead{Rigidity} & \colhead{Differential} &
\colhead{Rigidity} & \colhead{Differential} &
\colhead{Rigidity} & \colhead{Differential} &
\colhead{Rigidity} & \colhead{Differential} &
\colhead{Rigidity} & \colhead{Differential}
\\
\colhead{GV} & \colhead{Intensity\tablenotemark{a}} &
\colhead{GV} & \colhead{Intensity\tablenotemark{a}} &
\colhead{GV} & \colhead{Intensity\tablenotemark{a}} &
\colhead{GV} & \colhead{Intensity\tablenotemark{a}} &
\colhead{GV} & \colhead{Intensity\tablenotemark{a}} 
}
\startdata
4.333e-02&1.214e-05&2.814e-01&9.882e-04&2.473e+00&3.158e-02&7.145e+01&2.277e-05&2.914e+03&2.664e-10   \\
4.482e-02&1.571e-05&2.913e-01&1.066e-03&2.600e+00&3.169e-02&7.639e+01&1.867e-05&3.118e+03&2.148e-10   \\
4.636e-02&1.750e-05&3.016e-01&1.149e-03&2.736e+00&3.161e-02&8.167e+01&1.529e-05&3.336e+03&1.732e-10   \\
4.796e-02&1.894e-05&3.123e-01&1.239e-03&2.880e+00&3.133e-02&8.732e+01&1.253e-05&3.570e+03&1.394e-10   \\
4.961e-02&2.037e-05&3.233e-01&1.335e-03&3.033e+00&3.086e-02&9.337e+01&1.025e-05&3.820e+03&1.122e-10   \\
5.132e-02&2.189e-05&3.348e-01&1.439e-03&3.197e+00&3.019e-02&9.984e+01&8.395e-06&4.087e+03&9.020e-11   \\
5.308e-02&2.351e-05&3.466e-01&1.549e-03&3.371e+00&2.935e-02&1.067e+02&6.868e-06&4.373e+03&7.242e-11   \\
5.491e-02&2.526e-05&3.590e-01&1.667e-03&3.557e+00&2.833e-02&1.141e+02&5.616e-06&4.679e+03&5.808e-11   \\
5.680e-02&2.715e-05&3.718e-01&1.794e-03&3.755e+00&2.717e-02&1.221e+02&4.591e-06&5.007e+03&4.652e-11   \\
5.876e-02&2.918e-05&3.851e-01&1.929e-03&3.966e+00&2.589e-02&1.305e+02&3.752e-06&5.357e+03&3.720e-11   \\
6.078e-02&3.136e-05&3.988e-01&2.073e-03&4.192e+00&2.450e-02&1.396e+02&3.065e-06&5.732e+03&2.970e-11   \\
6.288e-02&3.372e-05&4.132e-01&2.227e-03&4.432e+00&2.302e-02&1.493e+02&2.503e-06&6.133e+03&2.367e-11   \\
6.504e-02&3.627e-05&4.280e-01&2.392e-03&4.689e+00&2.150e-02&1.597e+02&2.044e-06&6.562e+03&1.883e-11   \\
6.729e-02&3.902e-05&4.435e-01&2.568e-03&4.963e+00&1.994e-02&1.708e+02&1.668e-06&7.022e+03&1.495e-11   \\
6.960e-02&4.199e-05&4.596e-01&2.755e-03&5.256e+00&1.838e-02&1.827e+02&1.361e-06&7.513e+03&1.184e-11   \\
7.200e-02&4.519e-05&4.763e-01&2.955e-03&5.569e+00&1.683e-02&1.955e+02&1.110e-06&8.039e+03&9.354e-12   \\
7.448e-02&4.865e-05&4.937e-01&3.169e-03&5.903e+00&1.531e-02&2.091e+02&9.053e-07&8.602e+03&7.368e-12   \\
7.705e-02&5.239e-05&5.117e-01&3.397e-03&6.260e+00&1.384e-02&2.237e+02&7.380e-07&9.204e+03&5.785e-12   \\
7.971e-02&5.643e-05&5.306e-01&3.640e-03&6.641e+00&1.245e-02&2.392e+02&6.015e-07&9.848e+03&4.527e-12   \\
8.246e-02&6.080e-05&5.501e-01&3.900e-03&7.049e+00&1.112e-02&2.559e+02&4.900e-07&1.053e+04&3.529e-12   \\
8.530e-02&6.552e-05&5.705e-01&4.178e-03&7.484e+00&9.892e-03&2.738e+02&3.992e-07&1.127e+04&2.740e-12   \\
8.824e-02&7.064e-05&5.918e-01&4.475e-03&7.950e+00&8.746e-03&2.929e+02&3.250e-07&1.206e+04&2.117e-12   \\
9.128e-02&7.617e-05&6.139e-01&4.793e-03&8.448e+00&7.693e-03&3.133e+02&2.646e-07&1.290e+04&1.628e-12   \\
9.443e-02&8.215e-05&6.370e-01&5.133e-03&8.981e+00&6.734e-03&3.352e+02&2.154e-07&1.381e+04&1.245e-12   \\
9.769e-02&8.863e-05&6.611e-01&5.498e-03&9.550e+00&5.867e-03&3.586e+02&1.753e-07&1.477e+04&9.463e-13   \\
1.010e-01&9.564e-05&6.863e-01&5.890e-03&1.015e+01&5.089e-03&3.836e+02&1.426e-07&1.581e+04&7.143e-13   \\
1.045e-01&1.032e-04&7.126e-01&6.310e-03&1.081e+01&4.396e-03&4.104e+02&1.160e-07&1.692e+04&5.352e-13   \\
1.081e-01&1.114e-04&7.400e-01&6.762e-03&1.150e+01&3.783e-03&4.391e+02&9.438e-08&1.810e+04&3.977e-13   \\
1.118e-01&1.203e-04&7.687e-01&7.248e-03&1.225e+01&3.242e-03&4.698e+02&7.675e-08&1.937e+04&2.928e-13   \\
1.157e-01&1.300e-04&7.987e-01&7.771e-03&1.304e+01&2.770e-03&5.026e+02&6.240e-08&2.072e+04&2.134e-13   \\
1.197e-01&1.404e-04&8.301e-01&8.334e-03&1.390e+01&2.358e-03&5.377e+02&5.072e-08&2.218e+04&1.538e-13   \\
1.238e-01&1.518e-04&8.629e-01&8.941e-03&1.481e+01&2.002e-03&5.753e+02&4.122e-08&2.373e+04&1.094e-13   \\
1.281e-01&1.640e-04&8.974e-01&9.593e-03&1.578e+01&1.694e-03&6.155e+02&3.350e-08&2.539e+04&7.681e-14   \\
1.326e-01&1.773e-04&9.335e-01&1.029e-02&1.682e+01&1.430e-03&6.585e+02&2.722e-08&2.717e+04&5.305e-14   \\
1.372e-01&1.918e-04&9.713e-01&1.104e-02&1.794e+01&1.204e-03&7.046e+02&2.211e-08&2.907e+04&3.600e-14   \\
1.419e-01&2.074e-04&1.011e+00&1.185e-02&1.913e+01&1.012e-03&7.538e+02&1.796e-08&3.110e+04&2.395e-14   \\
1.468e-01&2.243e-04&1.052e+00&1.272e-02&2.041e+01&8.489e-04&8.065e+02&1.458e-08&3.328e+04&1.558e-14   \\
1.519e-01&2.426e-04&1.096e+00&1.364e-02&2.178e+01&7.103e-04&8.629e+02&1.184e-08&3.561e+04&9.879e-15   \\
1.572e-01&2.624e-04&1.142e+00&1.462e-02&2.324e+01&5.932e-04&9.233e+02&9.614e-09&3.810e+04&6.082e-15   \\
1.626e-01&2.839e-04&1.191e+00&1.565e-02&2.480e+01&4.945e-04&9.878e+02&7.803e-09&4.077e+04&3.620e-15   \\
1.682e-01&3.072e-04&1.242e+00&1.674e-02&2.648e+01&4.115e-04&1.056e+03&6.332e-09&4.363e+04&2.072e-15   \\
1.741e-01&3.323e-04&1.296e+00&1.787e-02&2.827e+01&3.419e-04&1.130e+03&5.137e-09&4.668e+04&1.132e-15   \\
1.801e-01&3.595e-04&1.353e+00&1.905e-02&3.018e+01&2.837e-04&1.209e+03&4.167e-09&4.995e+04&5.867e-16   \\
1.864e-01&3.890e-04&1.413e+00&2.026e-02&3.223e+01&2.350e-04&1.294e+03&3.379e-09&5.344e+04&2.847e-16   \\
1.929e-01&4.208e-04&1.476e+00&2.149e-02&3.442e+01&1.944e-04&1.385e+03&2.740e-09&5.719e+04&1.277e-16   \\
1.996e-01&4.552e-04&1.543e+00&2.273e-02&3.677e+01&1.607e-04&1.482e+03&2.221e-09&6.119e+04&5.200e-17   \\
2.065e-01&4.923e-04&1.613e+00&2.396e-02&3.928e+01&1.326e-04&1.585e+03&1.799e-09&6.547e+04&1.871e-17   \\
2.137e-01&5.324e-04&1.688e+00&2.517e-02&4.197e+01&1.094e-04&1.696e+03&1.457e-09&7.006e+04&5.734e-18   \\
2.212e-01&5.757e-04&1.767e+00&2.634e-02&4.484e+01&9.017e-05&1.815e+03&1.180e-09&7.496e+04&1.413e-18   \\
2.289e-01&6.224e-04&1.851e+00&2.744e-02&4.792e+01&7.423e-05&1.942e+03&9.558e-10&8.021e+04&2.553e-19   \\
2.369e-01&6.727e-04&1.940e+00&2.846e-02&5.121e+01&6.106e-05&2.078e+03&7.734e-10&8.582e+04&2.846e-20   \\
2.452e-01&7.269e-04&2.034e+00&2.937e-02&5.473e+01&5.019e-05&2.223e+03&6.256e-10&9.183e+04&1.351e-21   \\
2.538e-01&7.852e-04&2.134e+00&3.016e-02&5.849e+01&4.123e-05&2.379e+03&5.057e-10&9.826e+04&1.102e-23   \\
2.627e-01&8.481e-04&2.240e+00&3.080e-02&6.252e+01&3.385e-05&2.545e+03&4.087e-10&1.051e+05&4.036e-26   \\
2.719e-01&9.156e-04&2.353e+00&3.128e-02&6.683e+01&2.777e-05&2.723e+03&3.300e-10&\nodata & \nodata    
\enddata
\tablenotetext{a}{Differential Intensity units: (m$^2$ s sr GV)$^{-1}$.}
\end{deluxetable}

\end{document}